\documentclass[10pt, aps,prc,twocolumn,superscriptaddress,preprintnumbers,
amsmath, 
floatfix,
longbibliography,
nofootinbib
]{revtex4-1}
\usepackage[T1]{fontenc}
\usepackage[utf8x]{inputenc} 
\usepackage{adjustbox}          
\usepackage[caption=false]{subfig}
\usepackage{url}
\usepackage{color}
\usepackage{float}
\usepackage[pdftex,colorlinks=true, linkcolor = blue, citecolor=blue,urlcolor=blue, bookmarksnumbered=true, bookmarksopen=true]{hyperref}
\usepackage{longtable}
\usepackage{amsfonts}
\usepackage{dsfont}
\usepackage{wrapfig,bm} 
\usepackage[normalem]{ulem}
\usepackage{MnSymbol}
\usepackage{amsmath}
\usepackage{verbatim}
\DeclareMathOperator{\sinc}{sinc}

\newcommand{\beq}{\begin{equation}}
\newcommand{\eeq}{\end{equation}}
\newcommand{\bea}{\begin{eqnarray}}
\newcommand{\eea}{\end{eqnarray}}

\begin{document}
\title{A critical assessment of the current implementations of the Generator Coordinate Method for fission and heavy-on reactions}
  
\author{Aurel Bulgac}%
\affiliation{Department of Physics,%
University of Washington, Seattle, Washington 98195--1560, USA}

\date{\today}

\begin{abstract}

The generator coordinate method (GCM) was introduced in nuclear physics  by 
Wheeler and independently by Peierls  and their collaborators in 1950's and it is still one of the 
mostly used approximations for treating nuclear large amplitude collective motion (LACM). 
GCM was inspired by similar methods introduced 
in molecular and condensed matter physics  in the late 1920's, after the Schr\"odinger 
equation became the tool of choice to describe quantum phenomena.
The interest in the 1983 extension of GCM suggested by Reinhard, Cusson and Goeke, 
which includes the internal excitations (absent in the initial GCM formulation),
was revived in recent years. Unfortunately this newer version of time-dependent GCM (TDGCM) framework has  
flaws, which prevents it from describing correctly many anticipated features, 
in particular interference and entanglement, which can play an important role in fission and many-nucleon transfer reactions. 
I present here an alternative formulation, the enhanced GCM (eGCM), which 
is free of difficulties encountered in previous GCM implementations and which is relevant for fission and  
many-nucleon transfer in heavy-ion reactions, and which can be used in either static or time-dependent eGCM formulations.
In the eGCM framework the corresponding many-body waves functions have a much more complex structure and this framework 
is equivalent to a configuration interaction (CI) approach in the continuum for nuclear reactions. eGCM is aimed to be used in the microscopic 
description of heavy-ion reactions and fission in particular. The eGCM framework opens the possibility to evaluate  time-dependent lower 
bounds on the entropy in nuclear reactions and induced fission in particular, and thus on the evolution towards ``thermalization'' or equilibration. 
The eGCM framework is well suited to extract in a microscopic approach the induced fission cross sections, which are notoriously difficult to model. 

\end{abstract} 

\preprint{NT@UW-24-10}

\maketitle

\section{Assessment of various incarnations of GCM equations to non-equilibrium dissipative nuclear dynamics} \label{sec:I}

\begin{figure*}[htbp]
\includegraphics[width=\textwidth]{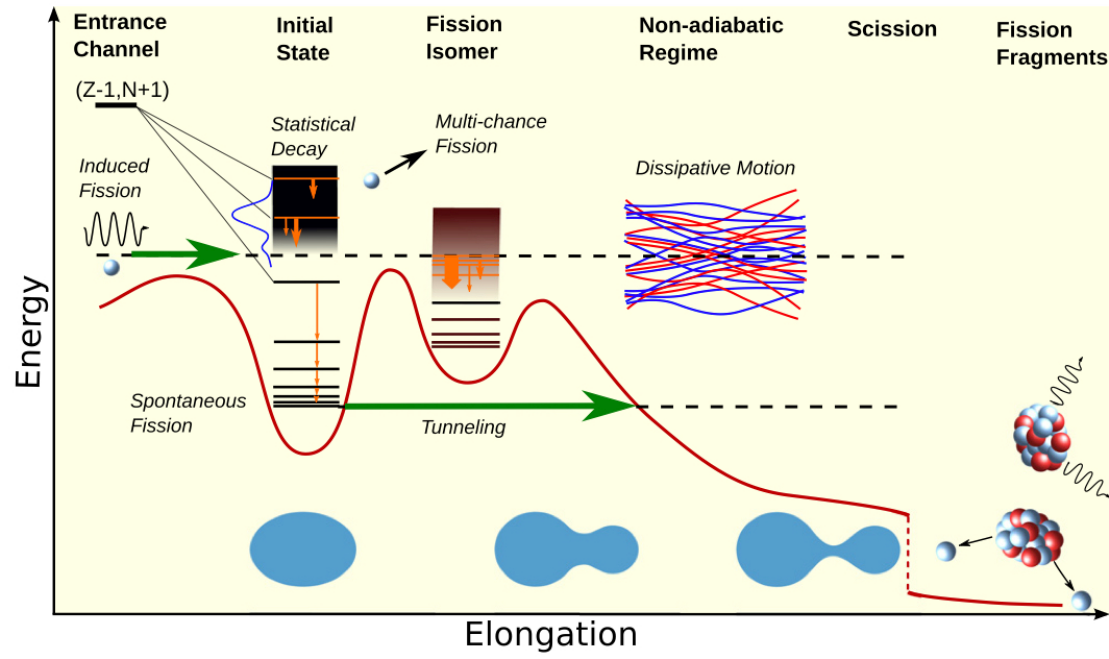}  
\caption{ \label{fig:00}  
 This open access figure from Refs.~\cite{Bender:2020,Schunck:2020},  courtesy of N. Schunck, vividly illustrates the complexity of the three main stages of induced fission. 
 In the initial state  and in the fission isomer stage  a fissioning nucleus spends approximately $10^{-15}$ seconds~\cite{Gonnenwein:2014}, while in the third stage, the non-adiabatic stage,  
 before scission the nucleus evolves during approximately $10^{-20}$ seconds~\cite{Gonnenwein:2014,Bulgac:2016,Bulgac:2019c,Bulgac:2020}, and the neck rupture 
 and the separation of the fissioning nucleus into FFs happens in about $10^{-21}$ seconds~\cite{Abdurrahman:2024}. 
 Subsequent processes, such as prompt neutron and light charge particle emission~\cite{Madland:2006}, 
 $\gamma$, and $\beta$ decay happen at much longer time scales~\cite{Gonnenwein:2014}.
 }
\end{figure*}  

The three stages of induced nuclear fission are all vividly illustrated in Fig.~\ref{fig:00}, where the vastly different time scales 
of the fission dynamics are very well separated in space and time. The efforts in the microscopic theory of fission  were correspondingly mainly focused in
time at different stages as well. The first  two stages, which at the time were unknowingly conflated in to a single stage, 
started with the brilliant insight due \textcite{Meitner:1939}, after which a large amount of energy is released as demonstrated by the experiement  of
\textcite{Frisch:1939}, followed by the theoretical studies of \textcite{Bohr:1939} and \textcite{Hill:1953} until early 1960s. 
The second stage, which came as a big surprise,  was defined by  the experimental discovery of the fission isomers in the late 1960s by \textcite{Polikanov:1968} 
and their theoretical justification with reasons to exist by \textcite{Strutinsky:1967} and others~\cite{Brack:1972,Bjornholm:1980,Vandenbosch:1973,Wagemans:1991}. 
This fast third stage, the time-dependent non-equilibrium and strongly dissipative fission dynamics was not treated in microscopic approaches until relatively recently,
see Ref.~\cite{Schunck:2016} for a theory review except in some phenomenological models, e.g. with classical statistical fluctuations, see Refs.~\cite{Frobrich:1998,Sierk:2017,Ivanyuk:2025} 
and many references therein, see Ref.~\cite{Schunck:2016} for a theory review.  The non-adiabatic and strongly dissipative character of the fission 
dynamics in the third stage  was firmly established theoretically only with the publication in 2016~\cite{Bulgac:2016}  and subsequent 
developments~\cite{Bulgac:2019c,Bulgac:2020,Bulgac:2021,Bulgac:2022b,Scamps:2023a,Abdurrahman:2024,Bulgac:2025a}.    
Prior to Ref.~\cite{Bulgac:2016} only a few (mostly) Time-Dependent Hartree-Fock attempts of describing induced fission of some transuranium elements were 
performed~\cite{Simenel:2014,Scamps:2015,Tanimura:2015,Goddard:2015, Goddard:2016}. These  
authors were apparently unaware of the crucial role played by pairing degrees of freedom (DoF) in fission~\cite{Bertsch:1997}, without which the 
FFs formation during the descent from saddle-to-scission is stalled. These authors started their simulations either with significant
excitation energy or large quadrupole deformation or at energies well below the ground state energy of 
the fissioning nucleus, thus excluding the dissipative non-adiabatic stage of the 
fission dynamics, which starts immediately at the top of the outer fission barrier~\cite{Bulgac:2016,Bulgac:2019c,Bulgac:2020}. 
Beyond the scission configuration the motion of the already formed FFs is dominated by 
the Coulomb repulsion between FFs alone~\cite{Frisch:1939} and the non-adiabaticity of the FFs formation dynamics is absent. 
An impinging low energy neutron leads to the formation of the compound nucleus in the ground state potential well,
 in a region with a very high  level density ${\cal O}(10^{5})/$MeV, which is comparable to the level density 
in the fission isomer stage. At the top of the outer fission barrier the compound nucleus is ``thermodynamically cold''~\cite{Bohr:1956,Wheeler:1963}, 
and at the scission point the energy level density increases to $\approx {\cal O}(10^{6})/$MeV or higher, as the nucleus at that stage 
reaches an intrinsic excitation energy of the order ${\cal O}(30)$ MeV, or even higher~\cite{Madland:2006,Bulgac:2019c}. Even though the 
``intrinsic temperature'' of the evolving nucleus is well above the average pairing gap, the role of the pairing correlations is still crucial and 
a lot of single-particle occupation probabilities are still changing at the same rate~\cite{Bulgac:2022,Bulgac:2023,Bulgac:2023c,Bulgac:2024a}. 
Pairing correlations are the only source of two-body collisions in a TDDFT framework.
The excitation energy of the FFs is eventually converted into the emission of several neutrons and many 
$\gamma$-rays~\cite{Gonnenwein:2014,Madland:2006}, with a small fraction of 
scission neutrons~\cite{Bohr:1939,Abdurrahman:2024}.

Starting near the outer fission barrier a nucleus evolves from a mixture of a ``few thermodynamically cold'' states 
into a system at a very high energy level density at scission configuration and beyond, with a high ``intrinsic temperature,'' 
related to the high intrinsic level density at that nuclear configuration. The total wave function of the nucleus at the scission 
configuration has a higher entropy, and thus is a linear combination of a significantly larger number of simple mean field states~\cite{Bulgac:2023,Bulgac:2023c,Bulgac:2024a}.

The GCM extension outlined in this work, the enhanced GCM (eGCM), is developed with the aim to describe this complex time evolution of the many-body wave function, 
which earlier GCM implementations did not incorporate. In all previous GCM implementations the many-body wave function of the nucleus, either 
in a static or time-dependent framework, was always a linear combination  of exactly the same number of  independent ``mean field'' configurations at all times, 
the GCM many-body wave function never increased its fragmentation into more (generalized) Slater determinants, 
thus the  complexity of the many-body wave function never increased during the nucleus evolution, or in other words the entropy of the system $S(t)$,
which is proportional to $\ln \rho(E_\text{int}(t))$, where $\rho(E_\text{int}(t))$ is the local level density at the instantaneous intrinsic excitation energy 
of the nucleus $E_\text{int}(t)$, never increased. The total energy of a fissioning nucleus uniquely separated into an intrinsic part and flow 
part for Galilean invariant energy density functionals, 
before and after scission~\cite{Engel:1975,Bender:2003,Bulgac:2013a,Bulgac:2018,Bulgac:2019c}
\begin{align}
&& E_\text{tot} \!\!=
E_\text{coll}(t)+E_\text{int}(t)
\equiv \int \!\!d{\bf r}\frac{ mn({\bf r},t){\bf v}^2({\bf r},t)}{2} \nonumber \\
&& +\int \!\!d{\bf r}\, {\cal E}\left (\tau({\bf r},t)-n({\bf r},t)m^2 {\bf v}^2({\bf r},t), n({\bf r},t),...\right ),\label{eq:etot}
 \end{align}
 where $n({\bf r},t)$ is the number density, $\tau({\bf r},t)$ is the kinetic density, and 
 ${\bf p}({\bf r},t)=mn({\bf r},t){\bf v}({\bf r},t)$ are linear 
 momentum and/or local collective/hydrodynamic velocity
 densities, ${\cal E}\left (\tau({\bf r},t)-n({\bf r},t)m^2 {\bf v}^2({\bf r},t), n({\bf r},t),...\right )$ is the energy density functional 
  and ellipses stand for various other densities.  
$\tfrac{{\bf p}({\bf r},t)}{n({\bf r},t)}$ is the position of the center of the local Fermi sphere in momentum space. 
 The first term in Eq.~\eqref{eq:etot} is the collective/hydrodynamic energy
 flow $E_\text{coll}$ and the second term is the intrinsic energy
 $E_\text{int}$ in the local rest frame.  For the sake of simplicity
 I have suppressed the spin and isospin DoF, even though they are included
 in all actual calculations.
 The collective energy $E_\text{coll}(t)$ is not vanishing only in 
 the presence of currents and vanishes exactly for stationary states. 

GCM was introduced in the 1950's by Wheeler and collaborators~\cite{Hill:1953,Griffin:1957} with the hope to eventually reach a microscopic treatment 
of nuclear fission as superposition of equilibrium nuclear liquid drop shapes, as it was initially envisioned by \textcite{Meitner:1939} and 
subsequently developed by \textcite{Bohr:1939}. 
In the 1960's it was established both theoretically and experimentally that nuclear fission is a more involved process, involving an additional intermediate second stage, 
the formation of fission shape isomers. Fission isomers are strongly deformed nuclear almost ``magic'' shapes, with large gaps in the single-particle spectrum, 
in a second well of  significantly larger deformations on the nuclear potential energy 
surface~\cite{Polikanov:1968,Strutinsky:1967,Brack:1972,Bjornholm:1980,Vandenbosch:1973,Wagemans:1991}. 
The transition from the second state to the third stage of the nucleus fission dynamics  is often treated theoretically 
using a transition state theory, a very old and familiar concept in chemical 
reactions~\cite{Vandenbosch:1973,Bohr:1956,Wheeler:1963}, when the excitation energy of the fissioning nucleus is 
close to the outer fission barrier, at which point the nucleus is thermodynamically ``cold.'' 

The heavy-ion reactions require also  a further development in theory, which prompted the next stage in the GCM development by \textcite{Reinhard:1983}. 
The properties of the fission fragments (FFs) are ultimately defined in the third stage 
of the fission process, when the nucleus passes past the outer fission barrier and the fission dynamics can be described correctly only as a time-dependent 
non-equilibrium dissipative process~\cite{Bulgac:2016,Bulgac:2019c,Bulgac:2020,Bender:2020}. The Time-Dependent Density Functional Theory (TDDFT) 
however can describe so far only average properties of the fission 
dynamics~\cite{Bulgac:2016,Bulgac:2019c,Bulgac:2020,Bulgac:2021,Bulgac:2022b,Scamps:2023a,Bulgac:2024,Bulgac:2025a}, which is main reason 
for the introduction of eGCM.  The Reinhard {\it et al.} GCM extension appear 
to be inaccurate, while attempting to reproduce experimental results for mass, charge, spins, excitation energies and the total kinetic energy distributions
of the FFs~\cite{Regnier:2019,Hasegawa:2020,Marevic:2023,Marevic:2024,Li:2023,Li:2024,Li:2025,Bjelcic:2026}. In different versions of 
Griffin and Wheeler GCM implementation the dissipative character of the saddle-to-scission fission dynamics is fully ignored, see 
Refs.~\cite{Schunck:2020,Schunck:2022} earlier references therein or treated phenomenologically. 

At its core GCM is a particular variant of the configuration interaction
(CI) general framework, albeit formulated in a non-orthogonal basis set of (generalized) Slater determinants, 
an approach discussed in atomic physics a long time ago, see Refs.~\cite{Lowdin:1955a,Lowdin:1955b,Lowdin:1955c,Lowdin:1956,Lowdin:1956a}.
The early attempts to describe molecular spectra by W. Heitler, F. London, F. Bloch, J. C. Slater, L. Pauling, and others 
used hybridized localized atomic orbitals, centered on atoms. Later F. Hund, R. S.  Mulliken, J. Lennard-Jones and others introduced 
delocalized molecular orbitals, which proved to be more flexible in practice.  
These ideas propagated further in condensed matter theory ~\cite{Bloch:1929,Mermin:1976} and lately in the high-temperature 
superconductivity~\cite{Gull:2022}.  These ideas also inspired J. A. Wheeler and his 
students~\cite{Hill:1953,Griffin:1957}  and also independently R. E. Peierls and 
collaborators~\cite{Peierls:1957,Yoccoz:1957,Peierls:1962} to introduce a similar description of nuclear 
LACM. These proposals were however a simplified version of the molecular and condensed matter frameworks, 
by using nucleon orbitals in a sequence of ground states of deformed mean fields  parametrized with quadrupole 
$Q_{20}$ and later also in addition with octupole deformations $Q_{30}$. The nuclear many-body wave 
function was represented as a linear superposition of these ``ground states'' of different shapes, 
within the lowest adiabatic Born-Oppenheimer approximation~\cite{Born:1927}. 
These are thermodynamically ``cold'' states as Bohr noted~\cite{Bohr:1956} in a different context,
In molecular physics non-adiabatic effects are due to the motion of the nuclei. In nuclear physics
the equivalent of motion of nuclei in molecules is played by the constrained nuclear shapes. In both atomic and nuclear physics 
the ``slow'' DoF are more often than not, not slow enough to justify invoking the adiabatic approximation.
No convincing arguments were presented in nuclear physics that the adiabatic approximation should be valid, 
particularly in LACM and in fission dynamics beyond the outer barrier in particular.
It was merely hoped that that would be the case. On the contrary,
it has been established by now that the descent of the fissioning nucleus from the top of 
the outer fission barrier to the scission configuration is, surprisingly, much slower 
than an adiabatic ``slide down the hill'' without friction~\cite{Bulgac:2019c}, which at first sight seems counter-intuitive also, 
while at the time this part of fission dynamics is also a
highly non-equilibrium and strongly dissipative process~\cite{Bulgac:2016,Bulgac:2019c,Bulgac:2020},
a conclusion which reached consensus among theorists~\cite{Bender:2020,Schunck:2020}.
These theoretical conclusions are also supported by the large total excitation energy of the FFs 
measured experimentally and their significantly lower total kinetic energy than the total energy released, which 
add up to approximately 200 MeV for induced fission of $^{235}$U, as \textcite{Meitner:1939} concluded in 
their seminal paper, where the term `` nuclear fission'' was also coined. 
These aspects are even more pronounced in non-relativistic heavy-ion reactions, were a significant fraction of 
the initial kinetic energy of the colliding partners is very efficiently converted into the 
excitation energy of the reaction products.

In nuclear literature the GCM was considered over the years in two flavors, 
with a time-independent ``generator wave function'' $f(Q)$ and with a 
time-dependent ``generator wave function'' $f(Q,t)$ respectively
\begin{align}
& \Psi(\xi_1\ldots\xi_A) = \sumint_{Q} f(Q) \Phi(\xi_1\ldots\xi_A|Q),\label{eq:GCM-s}\\
& \Psi(\xi_1\ldots\xi_A,t) = \sumint_{Q} f(Q,t) \Phi(\xi_1\ldots\xi_A|Q), \label{eq:GCM-t}
\end{align}
but in both cases with static (generalized) Slater determinants  $\Phi(\xi_1\ldots\xi_A|Q)
$~\cite{Ring:2004,Schunck:2016,Schunck:2020,Verriere:2020,Schunck:2022,Sadhukhan:2020,Bonche:1990}, generated as
a restricted set of constrained nuclear shape Hartree-Fock (HF) or Hartree-Fock-Bogoliubov (HFB) solutions. 
The set of ``shapes'' $Q$ in practice is always discreet and hence the use of the symbol $\sumint$.
I shall use the notation $|\Psi\rangle$ for GCM many-body wave functions and $|\Phi\rangle$ for HF or HFB many-body wave functions.
These choices of GCM states are neither well nor uniquely defined and consequently the ``sum'' over ``nuclear shapes'' 
parametrized by the multidimensional variable $Q$ lead to a uncontrolled approximation of the many-fermion wave function
$\Psi(\xi_1\ldots\xi_A)$ or $\Psi(\xi_1\ldots\xi_A,t)$. 
In a controlled approximation one would expect that the following  relation is satisfied
\begin{align}
&\sumint_{Q}\sumint_{Q'} \Phi(\xi_1\ldots\xi_A|Q){\cal N}^{-1}(Q,Q')\Phi^*(\xi_1'\ldots\xi_A'|Q')\nonumber \\
& = \delta (\xi_1-\xi_1'\ldots\xi_A-\xi_A'), \label{eq:complete}\\
& {\cal N}(Q,Q') = \sumint_{\xi_1\dots \xi_A} \!\!\!\!\! \Phi^*(\xi_1\ldots\xi_A|Q)\Phi(\xi_1\ldots\xi_A|Q'). \label{eq:norm}
\end{align}
Eq.~\eqref{eq:complete}  is never satisfied in practical GCM implementations and the accuracy of this approximation is not well 
understood~\cite{Bonche:1990,Dubray:2012,Verriere:2017,Carpentier:2024}. The main reason is due to the absence 
of a well defined small parameter and not a well understood meaning of a corresponding  energy cutoff $\Lambda$, 
in order to evaluate the magnitude of the neglected contributions. In constructing potential energy surfaces  (PES)
when defining the many-body wave functions $\Phi(\xi_1\ldots\xi_A|Q)$ one often runs into 
discontinuities, see Ref.~\cite{,Carpentier:2024} and some earlier references therein.  The continuity of the GCM basis many-body 
wave functions as a function of the ``collective'' $Q$, 
see Eqs.~(\ref{eq:GCM-s}, \ref{eq:GCM-t}), it is often a requirement in TDGCM implementations~\cite{Toledo-Piza:1978,Verriere:2020,Carpentier:2024}, 
which however it is not needed, as follows from Eq.~\eqref{eq:PsiQ} below and as I also discuss in Section~\ref{sec:III}.
\textcite{Carpentier:2024} solution to generate continuous PESs is to basically
enlarge the number of generator operators, from the typical set of quadrupole and octuples DoF.
At the same time, it is known for a long time from Thouless theorem~\cite{Thouless:1960,Ring:2004} that two (generalized) Slater
determinants can always be linked by a largely arbitrary and continuous unitary transformation.
These discontinuities  are most likely due to the projection of complicated energy surfaces into a lower
dimensional space, with 2 and rarely with 3 DoF, when ``folding'' of these surfaces occurs and which in 2D or in 3D 
appear as discontinuities, such as the umbilical catastrophes studied in Catastrophe 
Theory~\cite{Arnold:1972,Zahler:1977,Berry:1980,Arnold:1992,Gilmore:1993}. The appearance of discontinuities  implies that 
there is a need  for releasing unphysical constraints, which are not present anyway in a time-dependent many-body 
Schr\"odinger equation and either in a properly numerically implemented TDDFT framework.
 
In order to get an additional insight into the nature of the problem one tries to solve within GCM framework
I will make a detour by introducing the quantum many-body theory of the collective and intrinsic DoF, a process 
described a long time ago by \textcite{Feynman:1963}, from which either a classical ~\cite{Grange:1983,Weidenmuller:1984} or quantum 
Fokker-Planck equation~\cite{Bulgac:1996a,Bulgac:1998,Kusnezov:1999} for the ``collective DoF'' can be derived.
In the GCM framework one typically interprets the (generalized) Slater 
determinants labels $Q$ as a set of collective variables, thus practically introducing a poor man's version of the Feynman and Vernon's 
separation between the "intrinsic DoF  $\xi_{1}\ldots \xi_{A}$" and
the ``collective DoF $Q$.'' These collective DoF need to be ``requantized,'' within the 
GCM or TDGCM frameworks and in GCM this is achieved by adopting 
the Gaussian overlap approximation~\cite{Ring:2004,Schunck:2016,Verriere:2020,Schunck:2020,Schunck:2022} 
to the norm and Hamiltonian overlaps and ``deriving'' a typically second order partial differential equation 
Schr\"odinger-like  equation for these fictitious ``collective DoF $Q$.'' 

The non-orthogonality of the GCM functions $ \Phi(\xi_1\ldots\xi_A|Q)$, see Eq.~\eqref{eq:norm},  
is a rather serious technical problem in practice.
There exists however an appealing reformulation of this ``classic'' GCM formulation, in which this ``technical nuisance'' is eliminated.
Using the eigenvectors and the eigenvalues of the norm overlap ${\cal N}(Q,Q')$ one can introduce 
a new set of orthogonal generator many-body wave functions $|\overline{\Phi}_k\rangle$, see also Refs.~\cite{Lowdin:1955a,Lowdin:1955b,Lowdin:1955c,Ring:2004},
\begin{align}
& \sumint_{Q'} {\cal N}(Q,Q') g_k(Q') = \nu_kg_k(Q),\, \nu_k \geq 0, \\
&\sumint_Q g_k^*(Q)g_l(Q)=\delta_{kl},\label{eq:eig}\\
& |\overline{\Phi}_k\rangle =\nu_k^{-1/2}  \sumint_Q g_k(Q) |\Phi(Q)\rangle, \, \nu_k>0, \,\langle \overline{\Phi}_l|\overline{\Phi}_k\rangle =\delta_{kl}. \label{eq:PsiQ} 
\end{align}  
There is a more transparent way to define $|\overline{\Phi}_k\rangle$
\begin{align}
&\langle \xi_1\ldots\xi_A|\overline{\Phi} _k\rangle = \nu_k^{-1/2} \sumint_Q  \langle \xi_1\ldots\xi_A|\Phi |Q\rangle \langle Q| g_k\rangle ,\label{eq:qnu}\\ 
& \langle \xi_1\ldots\xi_A|\Phi |Q \rangle =\Phi(\xi_1\ldots\xi_A|Q),\,
 \langle Q| g_k\rangle =g_k(Q).
\end{align}
In all implementations of GCM in literature, only those $f_k(Q)$ with $\nu_k$ above a certain chosen value, 
which varies from one publication to another, are retained~\cite{Ring:2004,Schunck:2016,
Schunck:2020,Schunck:2022,Verriere:2020,Sadhukhan:2020,Bonche:1990}.
It is notable that the eigenfunctions of norm overlap corresponding to small eigenvalues $\nu_k\ll 1$ do not appear to be suppressed in Eq.~\eqref{eq:PsiQ}, 
see also Section~\ref{sec:III} for a few explicit examples.  In this new GCM basis $|\overline{\Phi}_k\rangle$ the equation for the many-body wave functions 
$|{\Psi}_n\rangle$ acquire the familiar expressions, as in any CI formulation of the many-body problem formulated in an orthogonal basis set
\begin{align}
& \!\!\!\!\!| \Psi_n \rangle = \sumint_k h_{n,k} |\overline{\Phi}_k\rangle,\quad \langle \overline{\Phi}_{n} |\overline{\Phi}_{m}\rangle =\delta_{nm}, \label{eq:Omega}\\
& \!\!\!\!\! \langle \overline{\Phi}_l | \hat{\rm H}|\Psi_n\rangle = E_n\langle\overline{\Phi}_l|\Psi_n\rangle, \, \sumint_k\langle \overline{\Phi}_l|\hat{\rm H}|\overline{\Phi}_k\rangle h_{n,k}=E_n h_{n,_k}. \label{eq:GCMT}
\end{align}
The orthogonality relation $\langle \overline{\Phi}_{n} |\overline{\Phi}_{m}\rangle =\delta_{nm}$ is a proof that all vectors $|\overline{\Phi}_{k}\rangle$ 
corresponding to $\nu_k>0$  are linearly independent and also might reveal that
the Hamiltonian overlap matrix has a non-trivial structure, with either an almost block-diagonal or only with a few relevant off-diagonals.
This has the implication that the selection of the "important"
configurations $|\Phi(Q)\rangle$ is not determined by whether some eigenvalues $\nu_k$ are large enough, 
thus implying that the corresponding eigenvectors $f_k(Q)$ can be neglected if $\nu_k$ are ``very small,'' but rather by the magnitude 
of the matrix elements $\langle \overline{\Phi}_l|\hat{\rm H}|\overline{\Phi}_k\rangle$ and the structure of this Hamiltonian hermitian matrix, as naturally expected 
in a matrix formulation of the Schr\"odinger equation in a reduced space of many-body wave functions, as in any CI framework.

In chemistry it has been known for decades that the adiabatic  
Born-Oppenheimer approximation is not accurate in many situations of interest in practice, see Ref.~\cite{Tully:1990} and 
many earlier references therein.  There exits also a number of  alternative somewhat related quantum many-body
frameworks, see Refs.~\cite{Requist:2016,Agostini:2020,Bulgac:2019d,Bulgac:2022} and earlier references therein. 
In the case of low energy nuclear LACM the nuclear shape should evolve in such a manner as 
to maintain the sphericity of the local Fermi momentum 
distribution~\cite{Bertsch:1980,Barranco:1988,Barranco:1990,Bertsch:1991,Bertsch:1997}, as otherwise 
the volume contributions to the energy of the nucleus  can become dominant in low-energy nuclear reactions. 
Only the Coulomb and surface isoscalar and isovector contributions to the total nucleus energy can vary considerably in fission in particular, 
in agreement with the brilliant insight of \textcite{Meitner:1939}. 
The energy contribution due to the emergence of pairing correlations  is always a small contribution, but however 
crucial~\cite{Bulgac:2016,Bulgac:2019c,Bulgac:2020,Bender:2020}, providing the essential ``lubricant'' for the 
compound nucleus to ``slide down the hill~\cite{Bulgac:2019c,Bulgac:2020}.''
In the Bethe-Weisz\"acker mass formula the odd-even correction term, which is related to pairing correlations, is about $3$ times 
smaller than the root mean square error of the binding energy 
of any nucleus with an atomic mass larger than $A\approx 80$~\cite{Ring:2004}. 
Pairing, however, plays the role of a  ``not quite perfect lubricant,'' but still allowing the nuclear shape to relatively 
easily hop form one shape to another in fission.
However, beyond the outer fission barrier the intrinsic excitation energy at a given nuclear shape is well above the lowest PES,
as the compound nucleus experiences ``quantum friction'' and by the time the compound nucleus reaches scission its temperature is quite high.  
As a result the LACM evolution of a fissioning compound nucleus is surprisingly significantly slower 
than in an adiabatic evolution, corresponding to a highly 
non-equilibrium dynamics~\cite{Bulgac:2016,Bulgac:2019c,Bulgac:2020,Bulgac:2025a}.

There were quite a number of attempts to ``fix'' some of GCM defficiencies 
by  \textcite{Peierls:1962}  and later in extensions of standard static GCM in
Refs.~\cite{Muther:1977,Goeke:1980,Bernard:2011,Chen:2017} in an approach where 
only a relatively small number of excited states was taken into account at each fixed value of the ``collective coordinate ${Q}$.'' 
GCM is one of the many-body quantum extensions of the simplest HF or HFB mean field frameworks. 
Apart from CI this list includes the many references Hartree-Fock method, random phase approximation, 
shell model, many-body perturbation approaches, and various algebraic methods and boson 
expansion approximations, coupled-cluster methods, {\it an initio} methods with chiral effective theory nucleon 
interactions, however typically for describing the properties of lowest energy nuclear states.
In each such extension arguments are made that only a restricted subset of (generalized) Slater determinants 
are sufficient to describe a particular set of nuclear excited states. 
Since LACM of a fissioning nucleus beyond the outer fission  barrier is a strongly dissipative 
non-equilibrium process~\cite{Bulgac:2016,Bulgac:2019c, Bulgac:2020,Bender:2020,Schunck:2020} 
the adiabaticity invoked by Wheeler and collaborators \cite{Hill:1953,Griffin:1957} is an unphysical assumption.
The argument brought forward in the case of spontaneous fission, which is to a large extent an 
under the barrier penetration process, is that in case of pairing effects with a noticeable large gap the adiabatic approximation is valid, 
at least in the case of even-even compound fissioning nuclei. 
This goes against the  solid theoretical arguments presented by \textcite{Caldeira:1983} and 
widely accepted in condensed matter physics, that the coupling to internal excitations, thus dissipation, 
leads to longer tunneling times.  These aspects are in qualitative agreement with 
experimental observations of the longer spontaneous life-times of  odd-mass 
and odd-odd nuclei~\cite{Vandenbosch:1973},  even though the fission barriers are quite similar to even-even nuclei, 
and also with recent theoretical findings~\cite{Bulgac:2024b,Bulgac:2025a} in case of neutron induced fission.

In 1983 \textcite{Reinhard:1983} suggested to replace the static fixed set of 
ground states  $ \Phi(\xi_1\ldots\xi_A|Q)$, through which a nucleus evolves during LACM, with the solutions of 
a time-dependent mean field problem. In their framework the total 
time-dependent nuclear wave function of the nucleus acquires a 
more complex structure, it is in general a linear combination over many
time-dependent Slater determinants   
\begin{align}
\Psi(\xi_1\ldots\xi_A,t)=\sumint_Q f(Q,t)\Phi(\xi_1\ldots\xi_A|Q, t). \label{eq:GCM_R}
\end{align} 
The hope was that within this prescription one may describe a dissipative non-equilibrium 
process such as nuclear fission or heavy-ion reactions.
It is crucial however to point \cite{Reinhard:1983} implicit and unjustifiable assumption made by \textcite{Reinhard:1983}, and currently used 
by a number of authors~\cite{Regnier:2019,Hasegawa:2020,Marevic:2023,Marevic:2024,Li:2023,Li:2024,Li:2025} that various
trajectories $\Phi(\xi_1\ldots\xi_A|Q,t)$ are started simultaneously, while they span a sufficiently large set of initial nuclear shapes, 
described by the shape (multidimensional) parameter $Q$.  The ``generator wave function'' $f(Q,t)$ 
is in this case a solution of the time-dependent Hill-Wheeler equation strictly within the span of $\Phi(\xi_1\ldots\xi_A|Q,t)$
\begin{align}
&i\hbar\partial_t \Phi(Q, t)= \hat{\rm H}_{MF}(t) \Phi(Q, t), \label{eq:mf0}\\
& i\hbar \sumint_{Q'} \langle \Phi(Q, t)| \Phi(Q',t)\rangle  \partial_t f(Q',t) \nonumber \\ 
&=  \sumint_{Q'} \langle \Phi(Q, t)|\hat{\rm H}-\hat{\rm H}_{MF}(t)| \Phi(Q',t)\rangle f(Q',t), \label{eq:GCM0}
\end{align}
where the nucleon coordinates $\xi_1\ldots\xi_A$ have been suppressed and the matrix 
elements 
\begin{align}
&{\cal N}(Q,Q'|t)=\langle \Phi(Q, t)|\Phi(Q',t)\rangle, \label{eq:GCMR} \\
&\langle \Phi(Q, t)|\hat{\rm H}-\hat{\rm H}_{MF}(t)| \Phi(Q',t)\rangle
\end{align}  
are evaluated by 
integrating over the coordinates $\xi_1\ldots\xi_A$. 
Above $\hat{\rm H}$ and $\hat{\rm H}_{MF}(t)$ stand for the many-body and mean field Hamiltonians respectively. 
In the TDDFT
extended to superfluid systems~\cite{Bulgac:2016,Bulgac:2019c,Bulgac:2020}, and as 
in any time-dependent mean field approach, the ``collective DoF $Q, Q'$'' in Eqs.~(\ref{eq:mf0}, \ref{eq:GCM0}) are 
merely labels for the parameters characterizing the initial nuclear shape, 
the equivalent of coordinates of nuclei in molecular physics. Notice also that one can replace 
$H_{MF}(t)$ with any other time translation operator, which one might found either convenient or more appropriate, 
see also discussion in Section~\ref{sec:III}.  


\section{Why the Reinhard, Cusson, and Goeke's prescription is physically unjustifiable?} \label{sec:II}

The essential deficiency in \textcite{Reinhard:1983}'s prescription, see Eq.\eqref{eq:GCM_R},  can be vividly illustrated  by the similarity between nuclear reactions 
and the Thomas Young two-slit (or even multi-grid) screen or the Mach-Zehnder experiments, with either light waves, electrons, atoms,  massive 
molecules, light from atoms~\cite{Taylor:1909,Merli:1976,Eichmann:1993,Aspden:2016,Arndt:1999,Grangier:1986,Aiello:2025,Fedoseev:2025,Zhang:2025}. 
In the case of the two-slit experiment $Q$ is the label 
of the slit where a specific ``classical''  trajectory $\Phi (\xi_1\ldots\xi_A|Q,t)$ was initiated at time $t=0$, since in all 
two-slit experiments the plane wave typically hits the two slits in a screen parallel with the front of the incident wave. 
$|\Psi(\xi_1\ldots\xi_A,t)|^2$, see Eq.~\eqref{eq:GCM_R}, is the total intensity 
of the ``combined beams'' described by $\Psi (\xi_1\ldots\xi_A,t)$,
 which hits the screen at position $(\xi_1\ldots\xi_A)$ at a later time $t>0$, which 
is a superposition of ``beams'' originating at different slits $Q$ and $Q'$. 
After the two-slit screen the two beams
travel however typically different times from the two 
slits labeled by $Q$ and $Q'$ to the particular position on the screen $(\xi_1\ldots\xi_A)$,
where fringes are observed. It is obvious then that Eq.~\eqref{eq:GCM_R} 
will not describe correctly the intensity  $|\Psi(\xi_1\ldots\xi_A,t)|^2$, since  by construction it ``constraints'' the two or more
independent ``light beams'' to travel with different speeds in order to arrive at any 
point on the screen $(\xi_1\ldots\xi_A)$ always at the same time $t$.  The time a particular ``classical trajectory'' $\Phi (\xi_1\ldots\xi_A|Q,t)$ travels 
from its initial ``slit'' to the final point $(\xi_1\ldots\xi_A)$ on the  ``screen'' or detector is different for different ``slits'' $Q$.
In induced fission for example, an incident low energy neutron beam, which is similar to the plane wave impinging on many slits,
excites the compound nucleus in its ground state potential 
well and within TDDFT different ``classical trajectories'' emerge, each labeled by the ``collective variable $Q$,'' 
and which reach the rim of the outer fission barrier  and eventually the FF detector at position $(\xi_1\ldots\xi_A)$, or,
analogously, a specific point on the ``screen,''  
where the mixing, or more correctly,  the beams interference occurs, with beams traveling up to the detector 
along different times~\cite{Bulgac:2016,Bulgac:2019c,Bulgac:2020,Bulgac:2025a,Bulgac:2024b}. 

One of the most extraordinary examples of such a type of mixing is the 
Hanbury Brown-Twiss (HBT) photon interference in astrophysics~\cite{Brown:1954,Brown:1956}, 
when two photons originating from two far sides of the Sirius star, 
which has a diameter $\approx 2.4\times 10^9$ m, hit a detector on Earth after traveling for vastly different times 
and being emitted at different times, with time arrival differences of the order of 10 sec., but generate a clear signal. 
In the HBT set-up two photon sources can be thought of as the two slits in the usual interference experiment 
and the two detectors as two spots on the screen~\cite{Boal:1990,Baym:1998}. In astronomy the separation between the sources is very large 
and the distance between the detectors is orders of magnitude much smaller, which is exactly opposite to 
the two-slit or many-slits experiments, where the separation between the diffraction gratings 
is much smaller than the separation between fringes on the screen.

The very large set of fission trajectories generated  in 
Refs.~\cite{Bulgac:2016,Bulgac:2019c,Bulgac:2020,Bulgac:2025a,Bulgac:2024b,Li:2024,Li:2025} and 
originating at different initial deformations  $Q\neq Q'$ correspond at a later time $t>0$ as a rule 
to vastly different shapes and very large separations between FFs in time and often in space as well
and in that case the Hamiltonian overlap $\langle \Phi(Q, t)|H| \Phi(Q',t)\rangle $ will most of 
the time be negligible, particularly beyond scission. On the other hand the overlap $\langle \Phi(Q, t)|H| \Phi(Q',t')\rangle $  
with $t\neq t'$ likely will not vanish, and particle transfer between such different mean 
field trajectories can occur, see discussions in Sections \ref{sec:III} and \ref{sec:V}.
\textcite{Boal:1990,Baym:1998} reviewed the HBT intensity interferometry, which can be used for studying 
correlations particles emitted in particle and nuclear collisions. 
Perhaps,  the intensity interferometry between FFs might be interesting to study, which would be somewhat more complex, since 
FFs are experimentally recorded only after neutrons and gammas are emitted  and possible ``ternary'' fission events. It is not clear to me yet, whether 
amplitude interferometry can be  experimentally contemplated yet in fission or heavy-ion reactions. Intensity interferometry 
however might provide additional information, apparently not yet exploited in fission studies, namely
\begin{align}
R_{Z_H,Z_L}=
\frac{\langle n_{Z_H,Z_L}\rangle}{\langle n_{Z_H}\rangle \langle n_{Z_L}\rangle}-1,
\end{align}
where $n_{Z_H,Z_L}$ are simultaneous and $n_{Z_H}, n_{Z_L}$ are  individual fission events recorded in FF detector 1 and 2 respectively. 
Since light charge particles (protons, deuterons, $\alpha$-particles, etc.) 
are emitted in fission with very low probability~\cite{Vandenbosch:1973}, the $R_{Z_H,Z_L}$, 
where the proton numbers in the heavy ($Z_L$) and light ($Z_L)$ FFs
and $Z_L+Z_L\leq Z$, where  $Z$  in the fissioning compound nucleus,  
could provide information, which could
possibly be further correlated with the total kinetic and possibly with the total excitation energies of the FFs. 
One can alternatively study $A_{H,L}$ instead.

The HBT ideas  have been used in nuclear physics~\cite{Boal:1990,Baym:1998} and 
applied for studying particle-particle correlations and similarly in cold atom systems~\cite{Jeltes:2007,Xiang:2025,Liu:2025,Yao:2025}.
The photon interference emitted at macroscopically different times was apparently first observed by \textcite{Taylor:1909} in a experiment 
with very "feeble light" and further confirmed in similar experiments with low intensity electrons~\cite{Merli:1976}, photons~\cite{Aspden:2016}, 
and even with buckyballs~\cite{Arndt:1999}. Thus there exist a vast amount of experimental data which confirms that
in a wave function there are very important contributions to the total wave function 
arising from contributions with different ''times.'' At a given spot on the screen electrons 
or photons hitting the screen at macroscopic times apart will show either a maximum or a minimum 
interference pattern~\cite{Merli:1976,Aspden:2016,Arndt:1999}. 
Nuclear reactions are more complicated, since the "interference" occurs between quantum objects 
with an internal structure. The multi-neutron transfer in heavy-ion collisions 
between the collision partners~\cite{Zagrebaev:2013,Sekizawa:2013,Sekizawa:2016,Sekizawa:2019,Simenel:2025,Pore:2024}
is a particularly interesting case, in which one might expect that an extended version of GCM can shed a lot of  light.
Obviously, at times $t\neq t'$  it is more likely than not that the Hamiltonian 
overlap  $\langle \Phi(Q,t)|H| \Phi(Q',t')\rangle $ is significant for $t\neq t'$, 
where $|\Phi(Q,t)\rangle$ and $|\Phi(Q',t')\rangle$ are two distinct mean field trajectories, initiated 
at different initial conditions $Q$ and $Q'$, which in this case represent impact parameters and orientations of the colliding 
deformed nuclei, and as a rule not necessarily started simultaneously.  

\begin{figure}[h]
\includegraphics[width=0.5\columnwidth]{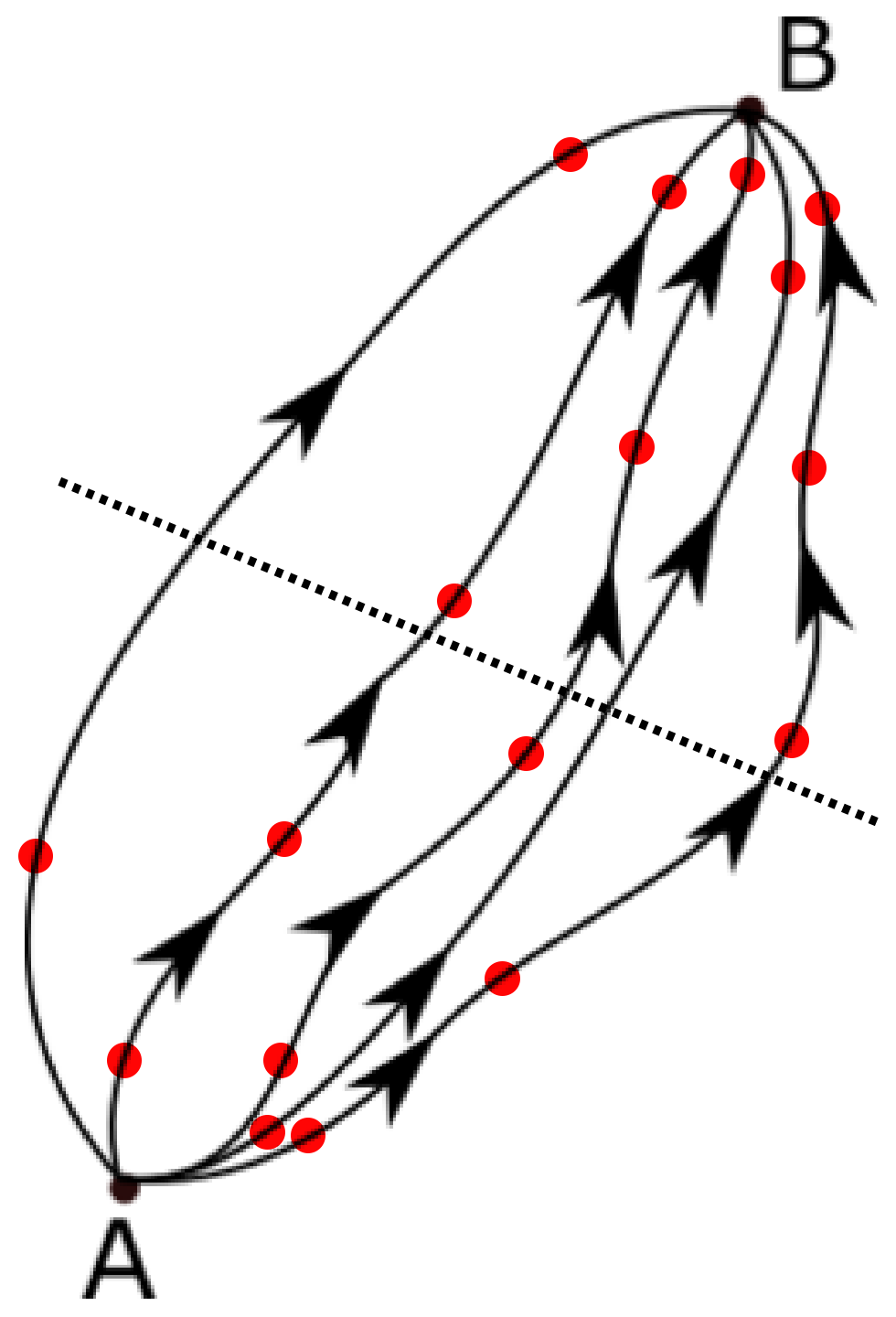}  
\caption{ \label{fig:0}  
 These are five of the infinitely many paths available in case of neutron induced fission from point A (the time of impinging neutron impact) 
 to point B (detector) at a later time.  This figure with added interaction times at random positions (red dots) is reproduced from Wikipedia 
 entry \emph{Path integral formulation}, is a pictorial representation of  a many-body 
 quantum propagator $K(\xi_1\ldots\xi_A,\tau| \xi'_1\ldots\xi'_A,\tau')$ from point ${\rm A}$ at time $\tau'$ to point ${\rm B}$ at time $\tau$.
 }
\end{figure}  

\textcite{Feynman:1948} presented a related argument  
a long time ago in quantum mechanics, see Fig.~\ref{fig:0}, when he introduced the ``sum'' over classical 
paths in order to construct a wave function or a propagator. 
In the path integral formulation the time-dependent many-body wave function is given by
\begin{align}
&\Psi(\xi_1\ldots\xi_A, t) =\exp\left( -\frac{i\hat{\rm H}t}{\hbar}\right )\Psi(\xi_1\ldots\xi_A, 0)= \\
&\int_{\xi_1'\ldots\xi_A'} K(\xi_1\ldots\xi_A,t| \xi'_1\ldots\xi'_A,0) \Psi(\xi_1\ldots\xi_A, 0),  \nonumber
 \end{align}
 where $K(\xi_1\ldots\xi_A,t| \xi'_1\ldots\xi'_A,t')$ is the many-body propagator, illustrated in Fig.~\ref{fig:0}
 
 The Reinhard {\it et al} TDGCM many-body wave-function  
 \begin{align}
 \Psi(\xi_1\ldots\xi_A, t)=\sumint_Qf(Q,t)\Phi(\xi_1\ldots\xi_A|Q,t)
 \end{align} 
 in  Eq.\eqref{eq:GCM_R} is a linear combination of mean field trajectories labeled by $Q$  
 along different paths started between A and B at $t=0$, 
 with the restriction the all trajectories end in B at exactly the same time $t$.
 
Consider the elastic collision of a light nucleus on a (very) heavy nucleus and take into account only the glancing trajectories, 
passing either mostly outside the target nucleus or through its surface mostly. 
In the detector one should observe the diffraction due to blocking of some of the trajectories due to the heavy target, in 
particular the equivalent of the famous Arago spot known in optics and predicted by Fresnel, the existence of which proved Newton 
and Poisson wrong (who both thought that this prediction was absurd) and Fresnel correct, and which was observed in the case of molecular beams 
as well by ~\textcite{Reisinger:2009} (see this reference for a few more riveting history details of this type of experiments, some performed as early as 1715).
The  ''Arago spot'' is related to the Ramsauer effect observed in nuclear collisions, in the measurements to the total cross sections, 
which according to the optical theorem is directly related to the imaginary part of the forward elastic scattering amplitude~\cite{LL3:1977}
\begin{align}
\sigma_{tot} = \frac{4\pi}{k} \mathfrak{Im}f(0).
\end{align} 
In particular, the total neutron cross sections experience strong oscillations as a function of 
neutron initial energy in the energy interval $E_n\in (1,1000)$ MeV, known as the Ramsauer 
effect~\cite{Peterson:1962,Bohr:1969}, first observed for electrons.

As known from the enormous experience in beyond mean field studies, the mixing of different 
configurations has nothing to do with their origin in any mean field or any other protocol chosen to generate them. 
In case of reactions the simplest choice 
for $|\Phi(Q,t)\rangle$ and $|\Phi(Q',t')\rangle$ are simple antisymmetrized products of Slater determinants of 
appropriate harmonic oscillator single-particle wave functions for each "fragment" $X_{1,2}$ at various 
positions along a classical Coulomb trajectory, which is close to the two-center shell 
model~\cite{Holzer:1969,Scharnweber:1970,Scharnweber:1971,Diaz-Torres:2005}, suggested more than six decades ago.
As discussed in Refs.~\cite{Peierls:1957,Peierls:1962,Kafker:2025}, in the case of an impinging nucleus 
on a static target two trajectories at two different times and close impact parameters will experience a strong mixing, see also {\it ii)} discussion below, 
even before the impact, at least in order to restore the CoM fluctuations of the projectile or of the target, see next section and Ref.~\cite{Kafker:2025}. 
The main deficiency of \textcite{Reinhard:1983}'s prescription and of its implementations used \emph{ad litteram} in recent 
studies~\cite{Regnier:2019,Hasegawa:2020,Marevic:2023,Marevic:2024,Li:2023,Li:2024,Li:2025} is that it departs from the basic argument 
in any beyond mean field method, that only the interaction matrix elements control whether 
two mean field configurations strongly mix or not. \textcite{Reinhard:1983} changed this natural beyond mean field 
condition with the requirement that two mean field trajectories $|\Phi(Q,t)\rangle$ and $|\Phi(Q',t)\rangle$
should reach the ``interaction region'' at the same time, while these trajectories are generated according to an arbitrary 
time-dependent mean field protocol, see Eqs.~(\ref{eq:GCM0}, \ref{eq:mf0}), and moreover, require as well 
that all trajectories should start simultaneously.    This is like asking for one person from each US contiguous states to 
leave their homes at the same time, drive at legal speed and arrive at exactly the same time in Washington DC., with no margin for errors.

There were a few recent attempts in 
literature~\cite{Regnier:2019,Hasegawa:2020,Marevic:2023,Marevic:2024,Li:2023,Li:2024,Li:2025}, 
where various authors  tried to strictly  implement the framework suggested by \textcite{Reinhard:1983}, 
but the results obtained so far have have not been particularly successful and in particular these 
failed  to describe data with a better accuracy than a quite wide variety of other phenomenological or simpler
approaches~\cite{Wilkins:1976,Brosa:1990,Lemaitre:2015,Ishizuka:2017,Sierk:2017,Albertsson:2020,
Ivanyuk:2024,Sadhukhan:2016,Sadhukhan:2017,Sadhukhan:2020}. 

\section{\lowercase{e}GCM equations} \label{sec:III}

The solution to the inconsistencies of the Reinhard {\it et al.} prescription  and implemented \emph{ad litteram} 
in recent studies~\cite{Regnier:2019,Hasegawa:2020,Marevic:2023,Marevic:2024,Li:2023,Li:2024,Li:2025} is rather simple.
One should consider using the ``running time'' along a TDDFT trajectory as an additional generator coordinate.
Instead of the norm and Hamiltonian overlap matrix elements in  Eq.~\eqref{eq:GCM0} proposed by \textcite{Reinhard:1983},
one should evaluate a set of  new generalized norm and the Hamiltonian overlap matrix elements
and solve the static enhanced GCM (eGCM) equations
\begin{align}
&\Psi(\xi_1\ldots\xi_A) = \sumint_{Q,\tau}f(Q,\tau)\Phi(\xi_1\ldots\xi_A|Q,\tau), \\
& E\sumint_{Q',\tau} \langle \Phi(Q,t) | \Phi(Q',\tau')\rangle  f(Q',\tau) \nonumber \\ 
&=  \sumint_{Q',\tau} \langle \Phi(Q,\tau)|\hat{\rm H}| \Phi(Q',\tau')\rangle f(Q',\tau') \label{eq:eGCM0},
\end{align}
using the variational approach with the eGCM many-body wave function $|\Psi(\xi_1\ldots\xi_A)$ defined in Eq.~\ref{eq:eGCM0}, 
where $| \Phi(Q,\tau)\rangle$ a time-dependent solution of the Eq.~\eqref{eq:mf0}, a procedure 
akin to a Feynman path integral representation for $|\Psi\rangle$, see discussion below.
Therefore, the states along different TDDFT trajectories are thus used as elements of a new eGCM basis set.
I use here the variables $\tau, \tau'$, as kindly suggested by P.G. Reinhard,  instead of the time variables $t, t'$ to make 
clear that $\tau, \tau'$ are merely labels for the ``generator time coordinates'', on par with $Q, Q'$ and not 
the real physical times in a time-dependent Schr\"odinger equation.
Thus they are merely labels for points along a chosen ``classical'' trajectory, as used for example in the Feynman path integral~\cite{Feynman:1948}, see Fig.~\ref{fig:0}.
As it is obvious from the eGCM equation Eq.~\eqref{eq:eGCM0}, the mixing of two states $|\Phi(Q,\tau)\rangle$ and $|\Phi(Q',\tau')\rangle$ 
is clearly allowed and they will mix  even if $\tau\neq \tau'$ and {\bf iff} $\langle\Phi(Q,\tau)|H|\Phi(Q',\tau')\rangle \neq 0$, as there is no 
conceivable physical argument to suggest that such transitions are or should be suppressed. In particular, 
such transitions are allowed even if the norm overlap matrix element identically vanishes $\langle\Phi(Q,\tau)|\Phi(Q',\tau')\rangle = 0$.
An obvious example is again the case of induced fission within TDDFT~\cite{Bulgac:2019c,Bulgac:2025a}, 
where quite often one encounters ``classical trajectories'' 
which turn back and forth and visits a very similar nuclear shape before the compound nucleus undergoes scission.

One can interpret the different paths in Fig.~\ref{fig:0} as light rays emanating from a point 
source and scattering from isolated atoms at different positions in space (red dots in the figure)
and thus each path having a different individual spatial length. This setup is somewhat similar
to a recent experiment with light scattering from cold atom clouds~\cite{Fedoseev:2025}.
Clearly in such a case Einstein's postulate in the theory of relativity  
that light propagates in vacuum (in between the scatterers from the source to the final screen) 
always with the same speed would come into conflict with the presumption that these rays are all emitted 
at the same time in A and all are recorded in B simultaneously at a later time.

One can introduce a slight generalization of the type of states 
described by Eq.~\eqref{eq:PsiQ} and obtain a more familiar 
type of the many body Schr\"odinger equation in an orthogonal 
basis set of eGCM many-body wave functions $|\widetilde{\Phi}_k\rangle$, see also Refs.~\cite{Lowdin:1955a,Lowdin:1955b,Lowdin:1955c}: 
\begin{align}
& \sumint_{Q',\tau'}  \langle \Phi(Q,\tau) | \Phi(Q',\tau')\rangle {\widetilde {g}}_k(Q',\tau') = \nu_k{\widetilde {g}}_k(Q,\tau),\label{eq:eig-eq} \\
&\sumint_{Q,\tau} {\widetilde {g}}_k^*(Q,\tau){\widetilde {g}}_l(Q,\tau)=\delta_{kl},  \quad {\rm only} \, \nu_k>0 \, {\rm are\, used}, \label{eq:eig-or}\\
& |{\widetilde {\Phi}}_k\rangle =\nu_k^{-1/2}  \sumint_{Q,\tau} {\widetilde {g}}_k(Q,\tau) | \Phi(Q,\tau)\rangle, \,  
\langle \widetilde{\Phi}_k |\widetilde{\Phi}_l\rangle =\delta_{kl}, \label{eq:PsiQe} \\
& |\Psi_n\rangle = \sumint_k h_{n,k} |{\widetilde {\Phi}}_k\rangle,\quad
\sumint_k\langle {\widetilde {\Phi}}_l|\hat{\rm H}|{\widetilde {\Phi}}_k\rangle h_{n,k}=E_n h_{n,l}, \label{eq:eGCMTef}\\
& \hat{\bf n}_{lk}(\xi,\xi')= 
\frac{ \langle {\widetilde {\Phi}}_l | \psi^\dagger(\xi)\psi(\xi') | {\widetilde {\Phi}}_k \rangle}
       { \langle {\widetilde {\Phi}}_l                                          | {\widetilde {\Phi}}_k \rangle}, 
& 
\end{align}
with hermitian one-body density $\hat{n}_{lk}(\xi,\xi')$ and Hamiltonian $\langle {\widetilde {\Phi}}_l |\hat{\rm H}|{\widetilde {\Phi}}_k\rangle $.
In standard GCM implementations there is an ongoing unsettled yet debate concerning the use of fractional powers for 
the one-body density operator $\hat{\bf n}$~\cite{Sheikh:2021} and earlier references therein, which sometimes can be resolved by using 
Pfaffians~\cite{Robledo:2009,Bertsch:2012} for the evaluation of the mean field wave functions overlaps $ \langle \Phi(Q,\tau) | \Phi(Q',\tau')\rangle$. 
The dimension of the set of orthogonal many-body wave functions ${|{\widetilde {\Phi}}_k\rangle  }$ 
is equal to the total number of non-vanishing eigenvalues $\nu_k$, which appears to be almost always equal 
to the total number of eGCM states $| \Phi(Q,\tau)\rangle$, see Section \ref{sec:V}. The orthogonalization process described 
above with Eq.~\eqref{eq:PsiQe} is mathematically equivalent to a Gramm-Schimdt procedure, which once again justifies the 
introduction of these states in the eGCM procedure.
The matrix $\hat{\rm M}$, see also Eq.~\eqref{eq:qnu},
\begin{align}
\langle k|\hat{\rm M}|Q,\tau\rangle = \nu_k^{-1/2}{\widetilde {g}}_k(Q,\tau)
\end{align}
 defines the 
transformation from the non-orthogonal GCM basis set $|\Phi(Q,\tau)\rangle$ to the orthogonal eGCM basis set $|\overline{\Phi}_k\rangle$.  
A time-dependent version of the eGCM equation with initial conditions for the expansion coefficients $h_k(t)$ is equally trivial
\begin{align}
&\Psi(\xi_1\ldots\xi_A,t)=\sumint_{k}h_{k}(t){\widetilde {\Phi}}_k(\xi_1\ldots\xi_A),\label{eq:PsiT}\\
& \sumint_k\langle {\widetilde {\Phi}}_l|\hat{\rm H}|{\widetilde {\Phi}}_k\rangle h_{k}(t)=i\hbar\partial_t h_{l}(t),\label{eq:eGCMT}\\
& h_k(0) = \int_{\xi_1\ldots\xi_A} {\widetilde {\Phi}}_k^*(\xi_1\ldots\xi_A) \Psi(\xi_1\ldots\xi_A,0),\label{eq:eGCMTT}
\end{align}
where now the time $t$ is the real physical time and assuming that the initial value of the many-body wave function 
$ \Psi(\xi_1\ldots\xi_A,0)$ is known.  Notice, that the typical GCM many-body wave functions do not have to be 
continuous functions of the ``collective'' variable $Q$, as integrals are always well defined even for discontinuous functions, 
at odds with restrictions often imposed in GCM literature~\cite{Toledo-Piza:1978,Verriere:2020,Carpentier:2024}.

The eGCM has a similar structure  to the Feynman path integral~\cite{Feynman:1948}, 
specifically when retardation effects are relevant. In the Feynman path integral
the lines represent  ``classical'' propagators of elementary particles, while in 
eGCM Eq.~\eqref{eq:eGCMT}, these lines 
represent  ``classical'' propagators of a many-body system described within TDDFT, where the
intrinsic dynamics is fully quantum and described by Eq.~\eqref{eq:mf0}.  

The eGCM equations 
Eqs.~(\ref{eq:eGCMTef}, \ref{eq:eGCMT})  are an explicit resolution of an issue raised by \textcite{Feynman:1948}, 
when discussing the role of field oscillators and the relativistic description of the wave function of two 
interacting particles $\psi(x_a, x_b; t)$. In that instance the behavior 
of particle $a$ at time time $t$ is specified by the behavior of the particle $b$ at an earlier time and vice versa. Feynman suggested a 
simple solution of this complicated multi-time problem, see Section 13 in Ref.~\cite{Feynman:1948}. 
\textcite{Feynman:1948} solves ``for the motion of the field oscillators before one integrates over 
the various variables $x_i$ ... which tries to condense the past history into a single state function.'' 
In this manner, the propagators for the oscillators appear in the path integral as a multiplicative of 
the action of the particles, taking into account the entire previous history of the oscillators for fixed particle positions. 
The dynamics and role of the ``oscillators'' is described within the eGCM framework by the TDDFT equations.
The TDDFT Eq.~\eqref{eq:mf0} for $|\Phi(Q,\tau)\rangle$ is solved before Eqs.~(\ref{eq:eGCMTef}, \ref{eq:eGCMT}) and thus the  earlier 
history is fully encapsulated in the many--body wave functions $|{\widetilde {\Phi}}_k\rangle$, following
thus identically the recipe suggested  by Feynman for `` field oscillators.'' 
Since both  the mean field or the  TDDFT equations can be rewritten as path integrals, it is now obvious that the entire
formalism described here is fully equivalent to a specific path integral formulation of the many-body Schr\"odinger equation.

I will exemplify now the eGCM framework with some very well-known cases when mixing TDDFT many-body wave functions at different 
times has been implicitly used in the past, without ever relating this procedure to a GCM procedure.

{\it i)}  Even though this aspect was never discussed in the initial formulation of the GCM 
approximation~\cite{Hill:1953,Griffin:1957} or even later, where a 1D sequence of nuclear shapes were considered, 
those shapes actually represent the fissioning nucleus at different times following a 
strictly adiabatic motion, as envisioned a long time ago by \textcite{Meitner:1939}, by 
implicitly using a time-dependent many-body wave function for a fission nucleus and implemented 
also implicitly by \textcite{Bohr:1939}. 
Thus mixing nuclear shapes at different times, should not come as a big surprise as this aspect was implicitly 
incorporated tacitly in GCM a long time ago~\cite{Hill:1953,Griffin:1957,Peierls:1957}.

{\it ii)} The \textcite{Peierls:1957} (P$\&$Y) prescription formulated however in the laboratory frame, 
as opposed to the center-of-mass (CoM) frame in which the translational invariance of a mean field wave function 
was restored by P$\&$Y,  is perhaps the simplest example demonstrating that eGCM trajectories mixing 
prescription is well defined.  Consider a free isolated nucleus in its CoM reference 
frame and additionally in the laboratory frame, in which the nucleus is moving with a constant speed. 
In order to restore the translational symmetry 
of the nucleus~\cite{Kafker:2025} one should mix in a  GCM framework the nucleus 
positions at different impact parameters and  at the same time  the nucleus positions along the direction of the motion. 
According to P$\&$Y the translational invariant many-body wave function in the nucleus 
CoM reference frame has the GCM-like expression
\begin{align}
\Psi({\bm r}_1\ldots{\bm r}_A) = {\cal N} \int \!\!\!d^3{\bm R} \Phi({\bm r}_1+{\bm R},...,{\bm r}_A+{\bm R}),  \label{eq:wf}
\end{align}
where ${\cal N}$ is a normalization factor, and where I suppressed the spin-isospin DoF.  One can now treat this nucleus 
as a moving nucleus with a constant velocity ${\bf v}= (0,0,v)$. By suppressing the irrelevant complex 
phase factor and after a simple change of integration variables to the moving reference frame  
${\bf R} \rightarrow \tilde{\bf R}(\tau)= (b_x,b_y,v\tau)$  Eq.~\eqref{eq:wf} becomes
\begin{align}
&\Psi({\bm r}_1\ldots{\bm r}_A)\equiv \nonumber \\
& v\!\!\!\int \!\!\!\! d^2{\bm b} \!\!\! \int \!\!\! d\tau \Phi({\bm r}_1+\tilde{\bm R}(\tau),...,{\bm r}_A+\tilde{\bm R}(\tau)), \label{eq:wf1}
\end{align}
where ${\bm b}=(b_x,b_y,0)$ is the ``impact parameter'' and the 
time-dependent mean field equations for the single-particle wave functions read
\begin{align}
& i\hbar\partial_\tau {\phi}_k({\bf r},\tau) = \frac{\hat{\bf p}^2}{2m}{\phi}_k({\bf r},\tau) +\hat{\rm U}_{MF}({\bf r},\tau){\phi}_k({\bf r},\tau),\\
& {\phi}_k({\bm r},\tau)= {\phi}_k({\bf r}+\tilde{\bm R}(\tau),\tau) \exp\left ( -i\frac{m{\bf v}^2\tau}{2\hbar}\right ), \\
&i\hbar \partial_\tau {\phi}_k({\bm r},\tau) = \frac{( \hat{\bm p}-m{\bf v})^2}{2m}{\phi}_k({\bm r},\tau) \nonumber \\
& \quad \quad \quad + \hat{\rm U}_{MF}({\bf r}+\tilde{\bm R}(t) ,t) {\phi}_k({\bm r},t),
\end{align}
where $\hat{\bf p}=-i\hbar{\bm \nabla}$ stands for the momentum operator.
These time-dependent mean field equations resemble the well-known 
in condensed matter Bloch's theorem~\cite{Bloch:1929,Mermin:1976}.
Eq.~\eqref{eq:wf} is the 
translational invariant wave function of a nucleus in its CoM frame. On the other hand,  the mathematically equivalent 
Eq.~\eqref{eq:wf1} is the translationally invariant many-body wave functions in a moving frame 
(apart from an irrelevant time-dependent phase factor),  
where a clear eGCM type of mixing occurs between many-body states at different times, unlike in 
the prescription advocated by \textcite{Reinhard:1983}, and exactly of the type of mixing I advocate here. 
In Eq.~\eqref{eq:wf1} ${\bf b}$ plays the role of $Q$.

{\it iii)} 
Consider an arbitrary time-dependent Schr\"odinger equation with an arbitrary time-independent Hamiltonian 
and for any number particles andwith an arbitrary initial condition
\begin{align}
&i\hbar\partial_t \Phi(t) = \hat{\rm H} \Phi(t), \quad \Phi(t) = \exp \left ( \frac{-i\hat{\rm H}t}{\hbar}\right) {\Phi}_0,
\end{align}
the equation can be used to find $\Phi(t)$ for any time $t\in (-\infty,\infty)$, and where ${\Phi}_0$ is not in general an energy eigenstate.
A solution of the static Schr\"odinger equation can be obtained by a simple Fourier transform
\begin{align}
&\sumint_{\tau'}\!\!\left [ \langle\Phi(\tau)|\hat{\rm H}|\Phi(\tau')\rangle -E\langle\Phi(\tau)|\Phi(\tau')\rangle \right ] g(\tau')=0, \\
& \Psi_E = \int_\tau\Phi(\tau)g(\tau), \quad \langle\Phi(\tau)|\Phi(\tau')\rangle\neq \delta(\tau-\tau'),\\
&\Psi_E \propto  \int_{-\infty}^{\infty}  \exp\left ( \frac{iE \tau}{\hbar} \right ) \Phi(\tau)d\tau, \, g(\tau) =  \exp\left ( \frac{iE \tau}{\hbar} \right ), 
\end{align}
thus restoring translational time invariance by mixing many-body wave functions at different times.
This is the time analog of the \textcite{Peierls:1957} GCM-like prescription, see Eq.~\eqref{eq:wf}, to find a state with a well defined
total linear wave vector ${\bf k}$  from an arbitrary many-body wave function $\Phi({\bf x})$ 
\begin{align}
&\Psi_{\bf k}({\bm r}_1\ldots{\bm r}_A) \propto \nonumber \\
&\int d^3{\bf R} \exp \left (-i{\bf k}\cdot{\bf R} - \frac{ i\hat {\bf P}\cdot {\bf R} }{\hbar}  \right) \Phi({\bm r}_1\ldots{\bm r}_A),\\
&\hat{\bf P}= \sum_1^A\hat{\bf p}_i, 
 \end{align}
 where spin-isospin coordinates were suppressed.  This is a standard projection technique
 used for restoring any quantum numbers for linear momentum, angular momentum, particle number, and parity
 from  many-body wave functions with ill-defined quantum numbers.

{\it iv)} There is a very simple implementation of eGCM framework using coherent states and similar inputs  in the case of one
particle, which can be shown to provide an exact solution for both static and time-dependent the Schr\"odinger equation. 
The orthogonal wave functions $\langle \xi_1\ldots\xi_A|\overline{\Phi}_k\rangle$ defined in 
Eq.~\eqref{eq:PsiQ}, in this case for one particle in 1D only can be chosen as Gaussians
\begin{align}
&\phi(x|a) = \frac{1}{ \pi^{1/4} \sigma^{1/2}  }\exp\left ( -\frac{ (x-a)^2 }{ 2\sigma^2 } \right ),\label{eq:Gauss}\\
&{\cal N}(a,b)=\int_{-\infty}^{\infty} \!\!dx \, \phi(x|a)\phi(x|b)= \exp\left ( -\frac{(a-b)^2)}{4\sigma^2}\right ),\\
&\int_{-\infty}^{\infty} db \exp\left ( -\frac{(a-b)^2)}{4\sigma^2}\right ) e^{ikb}  =  \nu_k e^{ika}, \label{eq:nu}\\
& \nu_k = 2\sigma\sqrt{\pi}\exp\left( -\frac{k^2\sigma^2}{2}\right), \quad g_k(a) = e^{ika}, \label{eq:nu_k}\\
&\overline{\Phi}_k(x) =\nu_k^{-1/2} \int_{-\infty}^\infty da \,\phi(x|a) e^{ika} = e^{ikx}. \label{eq:plane}
\end{align}
Note, one can use complex $a, b, \sigma$ with obvious generalization of these formulas.
In particular, one can consider a time-dependent form $a(t)$, with an arbitrary time-dependence in this case, when Eq.~\eqref{eq:nu} becomes
\begin{align}
\int_{-\infty}^{\infty} d\tau\,\dot{b}(\tau) \exp\left ( -\frac{(a-b(\tau))^2)}{4\sigma^2} \right ) e^{-ikb(\tau)}  =  \nu_k e^{ika}. \label{eq:nu1}
\end{align}
\textcite{Reinhard:1987} suggested another alternative set of single-particle wave functions instead of Gaussians,
\begin{align}
&\phi(x|a)= \frac{ \sqrt{\alpha/2} }{ \cosh[\alpha(x-a)] },\label{eq:cosh}\\
&{\cal N}(a,b)=\int_{-\infty}^\infty  dx\,  \phi(x|a)\phi(x|b)= \frac{\alpha(a-b)}{\sinh[\alpha(a-b)]},
\end{align}
which can be replaced with any other set of localized orbitals as well, see Ref.~\cite{Reinhard:1987} 
for a choice identical to the one suggested in Ref.~\cite{Peierls:1957}.
Since the set of wave functions defined in Eq.~(\ref{eq:plane}, \ref{eq:cosh}), or any other set off displaced orbitals
are complete, the accuracy of the solution  of the Schr\"odinger equation in these basis sets is under control, 
since there is a small parameter, the inverse of the energy cutoff 
\begin{align}
\Lambda = \varepsilon_c=\frac{\hbar^2k_c^2}{2m}.
\end{align}
 
This approach can be generalized to two particles
\begin{align}
\Phi(x,y|a,b)= \frac{\phi(x|a)\phi(y|b)-\phi(x|b)\phi(y|a)}{\sqrt{2 [1-\int_{-\infty}^\infty dx\phi(x|a)\phi(x|b)]} } ,
\end{align}
with the either discrete or continuous generator coordinate ${\bf Q}=(a,b)$ and eventually extended to any number of fermions.
For a choice of discrete set of discrete points $a_i$ the above formulas remain valid, 
with the exception of the actual numerical value for $\nu_k$ in Eq.~\eqref{eq:nu_k}. 
The use of such sets of single-particle wave functions has some similarities with the Discrete Variable Representation (DVR) 
of wave functions on a lattice, when one choses for $a, b, \,\dots$ a set of equally spaced discrete points~\cite{Bulgac:2013} 
and  the sinc functions
\begin{align}
&\phi(x|a) = \frac{\sin k_c(x-a)}{k_c(x-a)\sqrt{\pi}}=\frac{\sinc k_c(x-a)}{\sqrt{\pi}}, \\
&\!\!\int_{-\infty}^\infty \!\!\!dx \frac{\sinc k_c(x-a)}{\sqrt{\pi}}  \frac{\sinc k_c(x-b)}{\sqrt{\pi}}=\sinc k_c(a-b),
\end{align}
where $k_c=\pi/l$ is a cutoff wave vector and $l$ is the lattice constant. The DVR functions form an orthogonal set
in case of a discrete set of points $a-b = nl$ where $n$ is an integer.  Since in expansion in sinc functions 
is equivalent to a Fourier series~\cite{Bulgac:2013}, this becomes equivalent to an expansion in Gaussians
Ref.~\eqref{eq:Gauss} defined over a discrete finite set of equidistant points $a_i = il, \, i=1,\ldots N$.

The 3D extension of these formulas and the extension to many particle systems is equally valid.
Since plane waves  form a complete basis set coherent states of this type can be used in a GCM framework. 
This type of restricted single-particle wave functions has been introduced
for describing many fermion system in 3D
\begin{align}
\!\!\!\phi_i({\bf r},\tau) \propto \exp\left ( -\frac{({\bf r}-{\bf a}_i(\tau))\cdot \mathsf{T}_i(\tau)\cdot ({\bf r} -{\bf a}_i(\tau))}{2}\right ), \label{eq:coh}
\end{align}
up to a time-dependent normalization constant and with complex vectors ${\bf a}_i(\tau)$ 
and $\mathsf{T}_i(\tau)$ a symmetric complex $3\times 3$ matrix,
defining TD Slater determinants, 
and it is known in literature as the Fermionic Molecular 
Dynamics framework~\cite{Feldmeier:1990,Horiuchi:1991,Ono:1992,Ono:1993,Ono:2004}. 
Even though these kind of single-particle states are not orthogonal, the corresponding Slater determinants, overlaps of 
different such Slated determinants and corresponding are particle densities are straightforward to evaluate, see Ref.~\cite{Kafker:2025}.
Within eGCM apart from considering the  ``time generator coordinate $\tau$,'' as explained above, one has to add as 
well as generator coordinates also the values of the ``impact parameter'' ${\bf Q}_i= {\bf a}_i(0)-{\bf a}_i(0)\cdot \dot{\bf a}_i(0)/|\dot{\bf a}_i(0)|$. 
It is satisfying that in principle the eGCM framework with either a static displacement ${\bf a}$ or a time-dependent 
generator coordinate formulation with ${\bf a}(\tau)={\bf v}\tau$  is
thus equivalent to solving the Schr\"odinger equation  in the orthogonal basis $\overline{\Phi}_k({\bf r}_1\ldots{\bf r}_A)$, see Eq.\eqref{eq:plane},
with a controlled accuracy determined by the cutoff energy $\Lambda = \hbar^2k_c^2/2m$, where $\hbar k_c=\pi\hbar/l$ is 
the single-particle momentum cutoff.

The many-body Slater determinants constructed from pure harmonic oscillator 
single-particle wave functions  provide a qualitative description of nuclei~\cite{Bohr:1969,Ring:2004},
while at the same time the translation invariance can be restored after using the Peierls and Yoccoz projection 
procedure~\cite{Peierls:1957,Kafker:2025}.  Slater determinants constructed from single-particle wave functions 
used in Fermionic Molecular Dynamics, see Eq.~\eqref{eq:coh}, are sums of Slater determinants constructed from
pure harmonic oscillator single-particle wave functions. Using coherent states may provide thus a rather economic 
implementation of eGCM. In this respect GCM is analogous to the recipe used in shell model 
calculations~\cite{Johnson:2018}, where the use of a large set of harmonic single-particle wave 
functions,  used to construct many different Slater determinants, is the norm. 

As i have argued here, in case of LACM TDDFT can be used in eGCM to generate
a physically better choice set of many-body wave functions $\langle{\bm r}_1\ldots{\bm r}_A|\Phi(Q,\tau)\rangle$
in nuclear reactions, which is a significant extension over CI approaches. 
This set has the flexibility of being  ``complete enough,'' since a small parameter exists, the cutoff energy  $\Lambda$, and thus one can aim 
to describe entanglement and interference phenomena in nuclear reactions. This set basis set  
does not have to be directly related with the Hamiltonian $\hat{\rm H}$ used in Eq.~\eqref{eq:eGCMT}, as is the case also in 
CI calculations with molecular orbital-linear combination of atomic orbitals~\cite{Lowdin:1955a,Lowdin:1955b,Lowdin:1955c,Lowdin:1956,Lowdin:1956a}, 
and even simple combinations of Gaussian orbitals in chemistry~\cite{Popple:1972,Pople:1999}.  
From the discussion presented in previous sections, it is obvious that the eGCM framework, see  Eq.~\eqref{eq:eGCMT}, satisfies all the expected constraints for a 
beyond mean field extension of TDDFT equations, and these equations can now describe  
a time-dependent non-equilibrium dynamics of a very complex quantum many-body system of strongly 
interacting particles, this likely being a very competitive extension of existing frameworks described in   
Refs.~\cite{Tully:1990,Requist:2016,Agostini:2020,Bulgac:2019d,Bulgac:2022}.

\section{\lowercase{e}GCM feasibility} \label{sec:IV}

I provide now a concise evaluation of the eGCM computational 
feasibility and complexity and comparing eGCM with standard GCM and also with other CI type of studies 
of correlated nuclear many-body wave functions.
It is imperative at this stage, before embarking into implementing 
eGCM, to obtain an idea of the potential size of the problem one has to address, to develop an intuition 
concerning the meaning, size  and relevance of the parameter set $(Q, t)$
in order to asses  the feasibility of implementing eGCM. 
By allowing the time coordinate to become a generator coordinate the size of the generator 
coordinate space increases considerably. For a typical treatment of nuclear fission the sizes the corresponding norm and 
Hamiltonian overlap matrices is determined by the number of points ${\cal O}(1000)$ in the plane 
$Q_{20}, Q_{30}$~\cite{Regnier:2016a}. In analogy with a non-uniform diffraction grating with a large number of slits,  
one can associate  $Q$ with the label of a specific slit, as done currently in induced fission studies and
including not only the $(Q_{20}, Q_{30})$ position of the initial point on the rim of the outer barrier,  
but, if desired, also the Euler angles needed for the fissioning nucleus to perform an angular momentum projection.

There is a non-trivial aspect of the eGCM equations, particularly when there are two or more reaction fragments 
and one has to evaluate the Hamiltonian overlap $\langle \Phi(Q,\tau)|H| \Phi(Q',\tau')\rangle$. 
If the times $t\tau$ and $\tau'$ are not chosen carefully the fragments might be in too different regions 
of space and then the Hamiltonian overlap could become exponentially small and extreme 
technical/numerical skill is required to correctly account for such situations.  

\begin{figure}[h]
\includegraphics[width=0.81\columnwidth]{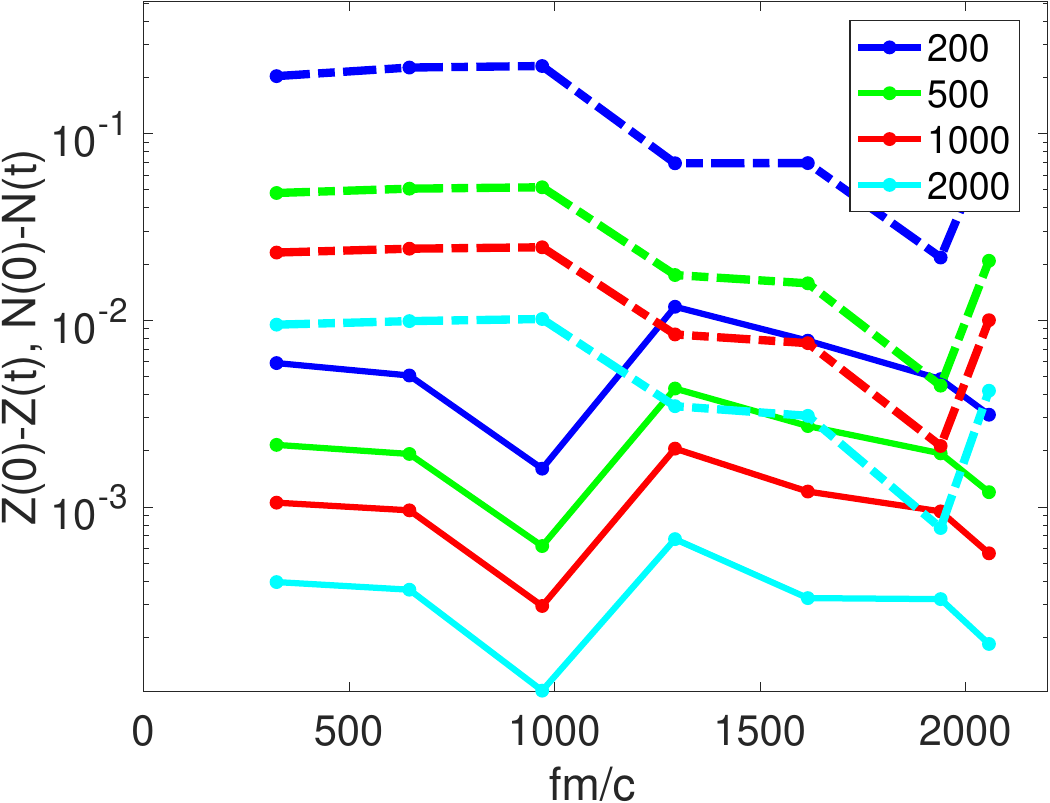}  
\caption{ \label{fig:DensityC}  
The proton (solid lines) and neutron (dashed lines) absolute error in particle numbers if only a reduced number of canonical quasi-particle 
wave functions $N_{cwfs}$ are used at different times, along an induced fission trajectory of $^{238}$U, to evaluate the total
number of nucleons, see Ref.~\cite{Bulgac:2024a} for more details.}
\end{figure}

The first main technical difficulty arises in evaluating the mixed densities for a single given set of values $(Q,\tau,Q',\tau')$ is 
the enormous number of needed quasiparticle wave functions and the corresponding mixed overlaps of the type
\begin{align}
\langle v^{Q,\tau}_k|v_l^{Q'\tau'}\rangle, \quad \langle u^{Q,\tau}_k|v_l^{Q'\tau'}\rangle,
\end{align}
involved in the construction of the densities $ n^{Q,\tau,Q',\tau'}(\xi,\zeta)$ and $\kappa^{Q,\tau,Q',\tau'}(\xi,\zeta)$. 
I will assume that one performs induced fission simulations in a typical ($64$ fm)$^3$ box, 
if angular momentum projection is considered. Otherwise
a $32^2\times64$ fm$^3$ simulation box is appropriate for axially symmetric even-even compound nuclei.
For evaluating the first and second spatial derivatives the use of Fast Fourier Transform  (FFT), which 
leads to machine precision and using powers of 2 is the best choice 
for these box sizes~\cite{Shi:2020, Bulgac:2011}. 
The lattice constant $l=1$ fm corresponds to a maximum momentum cutoff 
in one cartesian direction $p_{max} =\hbar \pi/l \approx $ 600  MeV/c, 
which is of the order of the maximum momentum cutoff considered in chiral effective field theory of nucleon interactions. 
It would be sufficient to  use a number $N_Q \leq 15$ for the set of quadrupole and octupole deformations $(Q_{20}, Q_{30})$
for an axially symmetric even-even compound nucleus along the rim of the outer fission barrier, as was done in 
Refs.~\cite{Bulgac:2016,Bulgac:2019c,Bulgac:2020}.  
Fortunately, as it was amply demonstrated recently, a quite accurate and economic solution already exists,  see again Fig. \ref{fig:DensityC} 
and Refs.~\cite{Scamps:2023a,Bulgac:2024a}, as it is sufficient to use 
not more than hopefully $N_{\rm{cwfs}} <  200\ldots 300$ canonical wave 
functions with the highest occupation probabilities,
 for the proton and neutron subsystems respectively. However, the construction of canonical wave functions 
 for each set of values $(Q,\tau)$ requires diagonalization of vey large matrices of sizes $2\times 32^2\times64$ in 
 case of fission of an even-even compound nucleus.

%
 
The second technical difficulty is in constructing the eGCM overlap matrices, either Eq.~\eqref{eq:eGCM0} 
or the physically more transparent Eq.~\eqref{eq:eGCMTef}.
In the case of fissioning even-even nuclei with axial symmetry one can build a rather simple angular momentum projection procedure
of the nucleus as only the Euler angles $\gamma$ and $\beta$ are needed and one can
use the icosahedral group with only 60 different angles to project on the total spin $J=0$ and 2, using the 
positions of the 60 carbon atoms in a $C_{60}$ buckyball for different orientations of the ``collective'' variables $(Q,Q')$. 
This approach is  orders of magnitude more economic that the usual angular momentum projecting 
techniques~\cite{Ring:2004,Lu:2022,Scamps:2023a}.
This leads to a conservative estimate of the dimension of the norm overlap matrices  (separately for protons and neutrons) 
$N_{eGCM} = N_{\rm{angles}} \times N_{times}\times N_Q = 60\times 60\times 15 =54,000$. 
If an angular momentum projection is not performed the dimension of norm overlap matrices is significantly smaller, 
namely $N_{eGCM}= N_{times}\times N_Q = 60\times 15 = 900$. 
 As it was recently shown in recent TDDFT simulation of strongly interacting fermions~\cite{Bulgac:2024a,Bulgac:2024}, the differences 
 between the number-projected and number-unprojected number 
 densities as a function of time at any point in space are at most at the level $0.5\%$ or less in the case of induced fission 
 treated in TDDFT with pairing correlations included, during the entire time-evolution from the top of the outer fission barrier to 
 complete FFs separation. These errors can be mitigated however if one remembers that their size is due to the Gibbs effect in FF.
 It is not surprising that TDDFT reproduces  the number projected number densities correctly by default.
Naturally, this new eGCM framework is equally applicable to other cases of nuclear LACM, in particular to collisions of heavy ions. 
Since the emergence of powerful supercomputers during the last two decades or so,  the numerical implementation of eGCM appears 
to be doable with many existing computer platforms.

In order to solve the eGCM equation Eq.~\eqref{eq:eGCMT} one needs to evaluate the 
mixed normal and anomalous densities~\cite{Ring:2004,Kafker:2025}
\begin{align}
&n^{Q,\tau,Q',\tau'}(\xi,\zeta) = \frac{\langle \Phi(Q,\tau)|\psi^\dagger(\zeta)\psi(\xi)|\Phi(Q',\tau')\rangle}{\langle \Phi(Q,\tau)|\Phi(Q',\tau')\rangle}, \label{eq:nn}\\
&\kappa^{Q,\tau,Q',\tau'}(\xi,\zeta) = \frac{\langle \Phi(Q,\tau)|\psi(\zeta)\psi(\xi)|\Phi(Q',\tau')\rangle}{\langle \Phi(Q,\tau)|\Phi(Q',\tau')\rangle},\label{eq:an}\\
&|\Phi(Q,\tau)\rangle = \prod _k [ u_k^*(\xi,\tau)\psi(\xi) +v^*_k(\xi,\tau)\psi^\dagger(\xi,t) ]|0\rangle, 
\end{align}
where $\xi$ stand for the spin-isospin and spatial nucleon coordinates,   
$\psi^\dagger(\xi), \psi(\xi)$ are creations and annihilation field operators for nucleons, $u_k(\xi,\tau), v_k(\xi,\tau)$ 
are TDDFT time-dependent quasiparticle wave functions~\cite{Bulgac:2013a,Bulgac:2019} 
with initial constrained ``shapes'' $Q$ at $\tau =0$, and $|0\rangle$ is the nucleon vacuum.  
In the case of generalized Slater determinants used in the case of induced fission
one has to include all allowed quasiparticle states in constructing accurate and 
physically meaningful time-dependent ``trajectories,'' see Ref. ~\cite{Bulgac:2024a}.
For a typical situation in a box $30^2\times60$ fm$^3$ and a lattice constant $l=1$ fm the number of 
3D+time partial differential equation required is a rather large number $2\times4\times32^2\times654$ = 524,288. 

The number of single-particle wave functions overlaps is thus $N_{sp-ovl}= N_{wfs} \times(N_{wfs} +1)$ (for both protons and neutrons)  
where $N_{wfs} = N_{\rm{cwfs}} \times N_{eGCM}$, which depending on whether one performs or not a 
total angular momentum projection is significant, however manageable with current computation facilities. Using for $N_{cwfs} =250$, 
see Fig.~\ref{fig:DensityC}, I obtain for $N_{wfs}=5\times10^{10}$ with no angular momentum projection 
using $N_{cwfs} = 250$. In the case of the reaction $^{48}$Ca+$^{208}$Pb I discuss 
in next section $N_{wfs}$ was of a similar magnitude. While eGCM appear to be quite massive numerical simulations, they 
compare favorably with current supercomputer resources, particularly when compared with state of the art 
shell-model calculations~\cite{Johnson:2018},  which deal with matrices of dimensions up 20 billion.

\section{Insights from from reactions and fission} \label{sec:V}

Multi-nucleon transfer reactions are the only way to create neutron rich nuclei at FRIB and other similar facilities and ultimately search for 
superheavy nuclei with charge $Z>218$~\cite{Gates:2024,Khuyagbaatar:2025,Mosat:2025} in particular and nuclei closer to the neutron drip line.
Microscopic methods used so far, like the Time Dependent Hartree-Fock (TDHF)~\cite{Sekizawa:2013,Sekizawa:2016,Simenel:2025} which has obvious limitations. 
Even the more complex coupled-channel~\cite{Hagino:1999,Hagino:2001,Hagino:2014} can be implemented for very low-energy collisions only.
M. Kafker and I~\cite{Kafker:2026a} implemented eGCM to the multi-nucleon transfer reaction  $^{48}$Ca + $^{208}$Pb, a reaction very close to 
those studied experimentally,  
and here I will present some of our preliminary results. These results will highlight the significant differences between 
using standard GCM, the GCM extension suggested by \textcite{Reinhard:1983} and implemented by a number of authors
recently~\cite{Regnier:2019,Hasegawa:2020,Marevic:2023,Marevic:2024,Li:2023,Li:2024,Li:2025}, and the need to add more complexity to GCM.

\begin{figure}[h]
\includegraphics[width=0.81\columnwidth]{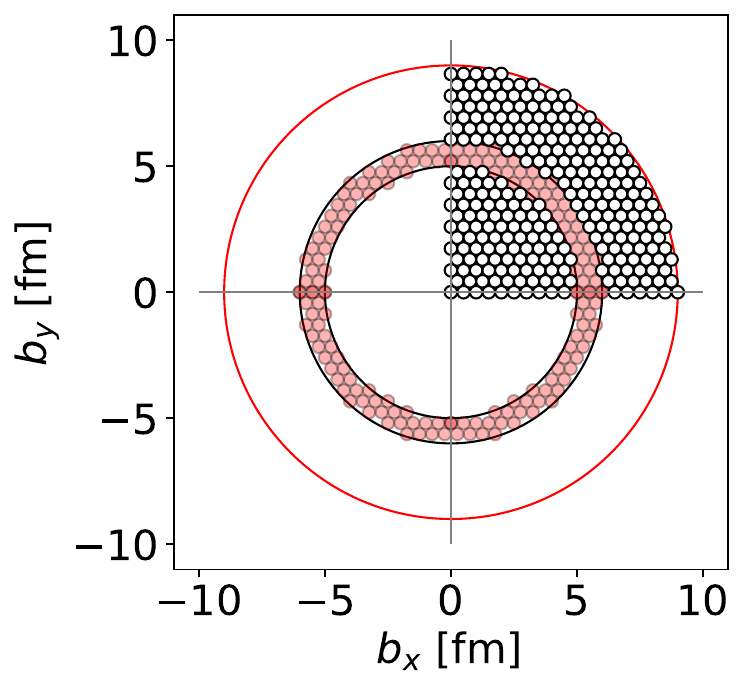}  
\caption{  \label{fig:ring}  
In the impact parameter plane $({\rm b}_x,{\rm b}_y,0)$ only the impact  parameter points with centers inside the orange ring, with $|{\bf b}|\approx 5-6$ fm, were
considered. For larger impact parameters the two nuclei do not form a neck and for a smaller parameters the formed ``compound nucleus'' 
fails to separate for very large simulations times. }
\end{figure}  

We performed Time-Dependent Hartree-Fock (TDHF) simulations of the reaction $^{48}$Ca + $^{208}$Pb, using the SeaLL1 
Energy Density Functional~\cite{Bulgac:2018} for a center-of-mass energy $E=235$ MeV near the Coulomb barrier  and 
with $N({\bf b})=152$ impact parameters corresponding to $|{\bf b}|\approx 5-6$ fm and azimuthal angles almost uniformly distributed  
in the range $\phi= (0\ldots 2\pi)$, see Fig.~\ref{fig:ring}. For larger values 
of the impact parameters the two colliding nuclei simply pass each other with negligible nucleon transfer. For smaller values 
of the impact parameter, $|{\bf b}|< 5$ fm, the ``formed compound nucleus'' fails to separate into fragments, unless followed up for very long times. 
By including all the angles $\phi= (0\ldots 2\pi)$ the axial symmetry of the wave function describing the reaction
$^{48}$Ca + $^{208}$Pb is restored. See also the remarks related to Eq.~\eqref{eq:wf1} concerning the restoration of translational symmetry.

The simulations are performed in a cubic box $(64 \,{\rm fm})^3$ with a lattice constant $l=1$ fm using the package LISE~\cite{Shi:2020}. 
The adopted value of the lattice constant $l=1$ corresponds to single-particle momenta up to $p_{max}=\pi\hbar/l\approx 600$ MeV/c.  
We evaluated all the norm overlaps $\langle \Phi({\bf b}, t)| \Phi({\bf b}',t)\rangle $ corresponding 
to the \textcite{Reinhard:1983} GCM extension and also the eGCM norm overlaps $\langle \Phi({\bf b}, \tau)| \Phi({\bf b},\tau')\rangle $.
These norm overlaps define orthogonal systems of GCM and eGCM many-body wave functions, see also 
Eqs.~(\ref{eq:norm}, \ref{eq:GCMR}, \ref{eq:eig-eq}, \ref{eq:eGCMTef}),
\begin{align}
& |\overline{\Phi}_k(t)\rangle =\nu_k^{-1/2}(t)  \sumint_{Q} \overline{g}_k(Q|t) | \Phi(Q,t)\rangle,\, {\rm for} \, {\rm GCM_R} \\ 
& |{\widetilde {\Phi}}_k\rangle =\nu_k^{-1/2}  \sumint_{Q,\tau} {\widetilde {g}}_k(Q,\tau) | \Phi(Q,\tau)\rangle, \, {\rm for} \,{\rm eGCM} 
\end{align}
where ${\rm GCM_R}$ stands for \textcite{Reinhard:1983} GCM extension.
These two sets of many-body wave functions  can be used for solving the ${\rm GCM_R}$ equations  and eGCM equations respectively. 
It is also of note that all eigenvalues $5\times 10^{-3}< \nu_k < 6$ in both cases.
 
Note, in the case of reactions, where there is an explicit time dependence, the standard GCM of \textcite{Griffin:1957} cannot be implemented, 
since during the collision the reaction partners acquire a significant amount of excitation energy, particle transfer between the 
reaction partners also occurs, which is the main reason why \textcite{Reinhard:1983} introduced their GCM extension.

The initial GCM equations suggested by \textcite{Griffin:1957}, the \textcite{Reinhard:1983} GCM extension  Eq.~\eqref{eq:GCM_R} 
and the present eGCM extension Eq.~\eqref{eq:eGCM0} are displayed below for clarity of comparison
\begin{align}
& \Psi(\xi_1\ldots\xi_A) = \sumint_{Q} f(Q) \Phi(\xi_1\ldots\xi_A|Q), \label{eq:GCM-s1}\\
&\Psi(\xi_1\ldots\xi_A,t)=\sumint_Q f(Q,t)\Phi(\xi_1\ldots\xi_A|Q, t), \label{eq:GCM_R1}\\
&\Psi(\xi_1\ldots\xi_A) = \sumint_{Q,\tau}f(Q,\tau)\Phi(\xi_1\ldots\xi_A|Q,\tau),  \label{eq:eGCM01}
\end{align}
where in the present case $Q$ stands for all the impact parameters. In Eq.~\eqref{eq:GCM-s1} one can also chose a time dependent function $f(Q,t)$ also.
While Eq.~ \eqref{eq:GCM_R1} mixes $N({\bf b})=152$ Slater determinants in the case of the reaction $^{48}$Ca + $^{208}$Pb, 
in eGCM implementation  the number of Slater determinants mixed is almost two orders of magnitude larger, namely $N({\bf b},\tau) = 8072$. 
As discussed in Section~\ref{sec:III} a time-dependent version of eGCM also exits and it is equally easy to implement, see Eqs.~(\ref{eq:PsiT}-\ref{eq:eGCMTT}).

\begin{figure}[h]
\includegraphics[width=0.81\columnwidth]{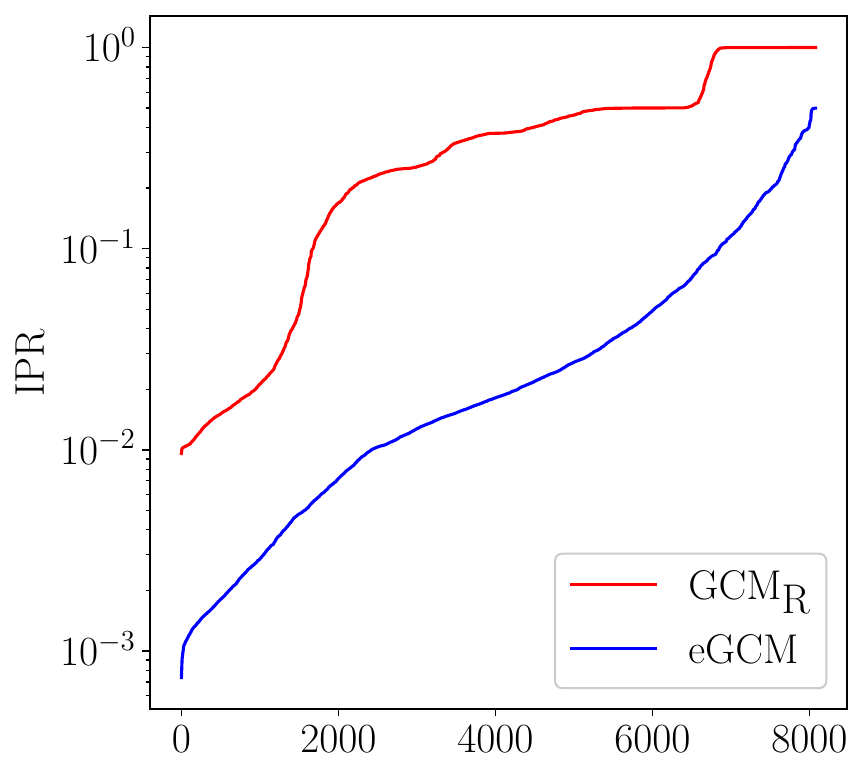}  
\caption{  \label{fig:IPR}  
The Inverse Participation Ratio (IPR) for all corresponding solutions to the GMR$_R$ 
and the present eGCM implementations for the corresponding eigenfunctions of the norm overlaps~Eqs. (\ref{eq:GCM00}, \ref{eq:GCM1}).
The eGCM IPR values are consistently well below the corresponding GCM$_R$ IPR values, 
thus with a significantly larger delocalization in eGCM,  corresponding to eGCM many-body wave functions with a 
significantly  more complex structure than the GCM$_R$ many-body wave functions. }
\end{figure}

The natural question arises, whether the eGCM the wave function in Eq.~\eqref{eq:eGCM01}, 
or its time-dependent counter part obtained by solving the corresponding initial value problem  described by Eqs.~(\ref{eq:PsiT}-\ref{eq:eGCMTT}),  is indeed a linear combination of
$N({\bf b},\tau) = 8072$ and not a linear combination between $N({\bf b})=152$ components only, as in Eqs.~(\ref{eq:GCM-s1}, \ref{eq:GCM_R1}).
In both cases one can construct an orthogonal set of  GCM wave functions. 
In the case of \textcite{Reinhard:1983} GCM extension $\Psi(\xi_1\ldots\xi_A,t)$, which  is 
at any time a linear combination of $N({\bf b})=152$ linearly independent wave functions at all times 
$$
{\overline{\Phi}}_k(t)\rangle =\nu_k^{-1/2}(t)  \sumint_{Q} \overline{g}_k(Q|t) | \Phi(Q,t)\rangle, \, {\rm same\, as\, Eq.~\eqref{eq:PsiQ}}.
$$
At the same time  the eGCM extension $\Psi(\xi_1\ldots\xi_A) $ is a linear combination of $N({\bf b},\tau) = 8072$ linearly independent wave functions
$$
|{\widetilde {\Phi}}_k(t)\rangle =\nu_k^{-1/2}(t)  \sumint_{Q\tau} {\widetilde {g}}_k(Q|\tau) | \Phi(Q,t)\rangle,\, {\rm same\, as\, Eq.~\eqref{eq:PsiQe}.}
$$
(As a side issue, with the \textcite{Reinhard:1983} wave functions $|\overline{\Phi}_k(t)\rangle$ one can also formulate a 
time-independent many-body equation with the initial conditions for the wave function Eq.~\eqref{eq:GCM_R1} as in eGCM, 
see Section~\ref{sec:III}.)
The eigenfunctions of the norm overlap matrix are the building blocks of the GCM many-body wave functions. 
We have evaluated the eigenvectors of the corresponding norm overlaps in \textcite{Reinhard:1983} GCM extension and in eGCM respectively, see Fig.~\ref{fig:IPR},
\begin{align}
\!\!\!\!\!& \sumint_{{\bf b}'} \langle \Phi({\bf b},t) | \Phi({\bf b}',t)\rangle \overline{g}_{k}({\bf b}',t) = \nu_{k}(t)\overline{g}_{k}({\bf b},t),  \label{eq:GCM00}\\
\!\!\!\!\!& \sumint_{{\bf b}',\tau'}  \langle \Phi({\bf b},\tau) | \Phi({\bf b}',\tau')\rangle g_k({\bf b}',\tau') = \nu_k{\widetilde {g}}_k({\bf b},\tau),  \label{eq:GCM1}
\end{align}
where in Eq.~\eqref{eq:GCM00} the time $t$ is the real physical time. 
In the case of \textcite{Reinhard:1983} implementation of the GCM the overlaps, by selecting only the block diagonal part 
of the entire eGCM overlap matrix $ \langle \Phi({\bf b},\tau) | \Phi({\bf b}',\tau')\rangle $ corresponding to equal times $\tau=\tau'$ 
blocks one obtains that the corresponding eigenvectors are non-vanishing only in each block, thus identical to GCM$_R$ eigenvectors, namely
\begin{align}
\overline{g}_k^*(Q,t)\overline{g}_l(Q',t') \equiv 0 \, \, {\rm if} \, t\neq t' ,\, \overline{g}_k({\bf b},t)\equiv \widetilde{g}_k({\bf b},t). \label{eq:Rgcm}
\end{align}  

Now one can evaluated the inverse participation ratio (IPR) in these two cases
\begin{align}
&{\rm IPR}_{GCM_R,k}(t)  = \sumint_{{\bf b}}        | \overline{               g } _k({\bf b},t     )|^4 \geq  \frac{1}{N({\bf b})      } = \frac{1}{152} , \\
&{\rm IPR}_{eGCM, k}       = \sumint_{{\bf b},\tau} | {\widetilde {g}} _k({\bf b},\tau)|^4 \geq \frac{1}{N({\bf r},\tau)}    = \frac{1}{8072}, \\
& {\rm IPR}_{GCM_R,k}(t)\gg {\rm IPR}_{eGCM, k} , \quad {\rm see} \quad {\rm Fig.~\ref{fig:IPR}}. 
\end{align}
The static and time-dependent GCM and GCM$_R$ many-body wave functions are at all times a superposition of at most 
$N(Q)$ mean field states, unlike the eGCM many-body wave functions which are a superposition of $N(Q,t)\gg N(Q)$ states.
The IPR characterizes the degree of delocalization of a wave function, which reaches the minimum value 
when the wave function is fully delocalized, while for a localized state ${\rm IPR}\equiv 1.$ A strong delocalization of 
the basis set of the many-body wave functions is the right ingredient required  to achieve microscopically derived mass, charge, and 
kinetic energy distributions of fission fragments in agreement with experiment. 
An eGCM many-body wave function has a significantly more complex structure than either the standard GCM of \textcite{Griffin:1957} 
or the GCM extension of \textcite{Reinhard:1983}, as it is obvious from the corresponding values of ${\rm IPR}$.

The massive difference between the delocalization of the many-body wave function within standard GCM~\cite{Griffin:1957}, see Eq.~\eqref{eq:GCM-s1},  
\textcite{Reinhard:1983} extension of GCM, see Eq.~\eqref{eq:GCM_R1},  and eGCM, see Eq.~\eqref{eq:eGCM01}, 
is the result of the  mixing of the mean field trajectories with different times. As mentioned 
before, the evaluations performed in Refs.~\cite{Regnier:2019,Hasegawa:2020,Marevic:2023,Marevic:2024,Li:2023,Li:2024,Li:2025,Bjelcik:2025} 
for the  various observables are narrower than the experimental measured ones, which is a
telling sign that the microscopic many-body wave functions evaluated in various suggested so far incarnations of GCM are not ``delocalized'' enough. 
Even though not as spectacular, the eGCM delocalization mechanism described here has similarities with the Cooper's formation mechanism 
of pairs~\cite{Cooper:1956}, when an insulator becomes a superconductor, when the localized wave function of an electron pair is 
strongly delocalized over large distances. 
In the case of  the reaction $^{48}$Ca + $^{208}$Pb, and initial wave function localized in TDHF at the impact parameter 
${\bf b} =(0,{\rm b},0)$ it is strongly delocalized to fully fill in the 2D entire ring highlighted in Fig.~\ref{fig:ring} and in addition with  a further 
delocalization  due to the time-mixing in eGCM.  As discussed in relation to Eq.~\eqref{eq:wf1} the axial and 
translational symmetry along the $z$-axis is restored before that two nuclei come into contact. And since TDDFT conserve symmetries, the axial symmetry 
of the total many-body scattering wave function axial symmetry is thus conserved at all times, as theoretically expected, 
with the bonus that center-of-mass energy correction in the initial state is also accounted for, see Eq.~\eqref{eq:wf1} and Ref.~\cite{Kafker:2025}. 
For a full annalysis of they reaction within eGCM see our recent results~\cite{Kafker:2026a}, where a number of new aspects have been 
introduced and discussed as well. The reader should notice also that the definition of IPR in Ref.~\cite{Kafker:2026a} is the inverse 
of the IPR defined here, which is a more informative characterization of the complexity of the eGCM many-body wave function. The results reported in Ref.~\cite{Kafker:2026a} also correspond to a significantly finer time discretization $\Delta \tau = 6.4$ fm/c, which leads to a significantly larger 
$N({\bf b},\tau) = 39630$ than the value used to obtain the eGCM IPR values  in Fig.~\ref{fig:IPR}. where $N({\bf b},\tau)= 8072$.

\section{Taking Stock and Conclusions} \label{sec:VI}

While \textcite{Reinhard:1983} suggested their GCM extension in 1983  in order to describe heavy-ion reactions, it is remarkable that nowadays 
this approach is basically ignored by many researchers  interested in creating nuclei close to the neutron drip line or superheavy 
elements~\cite{Zagrebaev:2008,Zagrebaev:2008a,Oganessian:2006,Rachkov:2014,Hinde:2021,Godbey_2025}, while at the same time recognizing 
the paucity of reliable microscopic models and the drawbacks of many used phenomenological approaches and the difficulties encountered by experimentalists 
to interpret  the experimental data~\cite{Hinde:2021,Godbey_2025,Cook:2023} using phenomenological models. Even though the creation of superheavy elements is a very complex quantum many-body problem 
involving vastly different time scales, at least some of these stages can nowadays be treated microscopically, as is the case of the last stage of induced fission and of 
the entire $^{48}$Ca+$^{208}$Pb reaction and formation of the primary reaction products, 
a prototype of the reactions used in superheavy element studies~\cite{Godbey_2025}. The discussion in 
Section~\ref{sec:V} proves that the TDHF is not really ``a computationally expensive method'' on leadership class 
supercomputers, as stated in  Ref.~\cite{Godbey_2025}.  The generation of 
the hundreds of TDDFT trajectories for fission are essentially routine~\cite{Bulgac:2025a,Bjelcik:2025} 
and the implementation of eGCM is feasible, see Section~\ref{sec:IV}. 

 The complexity of a many-body wave function, 
in particular the complexity of a (TD)DFT many-body wave function, is defined 
by the minimal number of Slater determinants or components needed to exactly represent it, namely~\cite{Bulgac:2023}
\begin{align}
\frac{N_{sp}!}{N! (N_{sp}-N)!}\approx \left ( \frac{N_{sp}}{N}\right )^N \gg 1,
\end{align}
where $N$ is the number of fermions  and $N_{sp}$ the number of single-particle fermionic states states, which typically is (much) larger than $N$, 
depending on the particular mean field implementation. Another measure is the lower bound to the time-dependent entropy discussed below.
The full many-body function of a nucleus is much more complex than a mean field many-body wave function, since in practice 
with controlled  accuracy one can ignore the breaking of various symmetries and the quantum fluctuations of various components of the mean field.
For a large number of purposes such ``subtleties'' can be ignored, e.g. if the ground state energies are not needed with an accuracy less than 2-3 MeV, 
as it is quite often the case in many mean field studies.

The GCM many-body wave functions have an additional degree of complexity, as they are linear combinations 
of many (TD)DFT many-body wave functions, each with a very large complexity as discussed in the above paragraph and in Ref.~\cite{Bulgac:2023}.  
The most important aspect of eGCM framework when compared to previous GCM formulations 
it is  the vastly different complexity of the emerging many-body wave functions. The minimal number of independent components 
of the eGCM many-body wave functions is two or more orders of magnitude larger than the corresponding number 
for GCM many-body wave functions, as it is obvious from the corresponding representations of the many-body wave functions, since $N(Q,\tau)\gg N(Q)$, 
\begin{align}
\!\!\!\!\!&\Psi_{GCM_G}(\xi_1\ldots\xi_A,t) = \sumint_{Q}f(Q,t)\Phi(\xi_1\ldots\xi_A|Q), \label{eq:GCM_G2}\\
\!\!\!\!\!&\Psi_{GCM_R}(\xi_1\ldots\xi_A,t) = \sumint_{Q}f(Q,t)\Phi(\xi_1\ldots\xi_A|Q,t),\label{eq:GCM_R2}\\
\!\!\!\!\!&\Psi_{eGCM}(\xi_1\ldots\xi_A) = \sumint_{Q,\tau}f(Q,\tau)\Phi(\xi_1\ldots\xi_A|Q,\tau),\label{eq:eGCM2}
 \end{align}
where $GCM_G$ and $GCM_R$ stand for \textcite{Griffin:1957} and \textcite{Reinhard:1983} GCM many-body wave functions formulations respectively.
Both $GCM_G$ and $GCM_R$ many-body wave functions have a similar complexity, they are linear combinations of exactly 
the same number of independent mean field many-body wave functions at all times, namely the number of GCM 
configurations selected $N(Q)$ at time $t=0$. This is the reason why both version of the GCM many-body wave 
functions failed to describe accurately distributions of mass, charge, excited energies of fission fragment and of their total kinetic energies so far. 
These GCM wave functions lack enough complexity. On the other hand the eGCM many-body wave functions have 
a complexity  almost two orders of magnitude larger than either  GCM$_G$ or  GCM$_R$ implementations. Th eGCM many-body wave 
function is practically a Feynman path integral~\cite{Feynman:1948}  over variables $(Q,\tau)$.

The standard GCM$_G$  and GCM$_R$ extensions
both represent the many-body wave function ``in the detector'' or at scission (depending on particular implementations) as a combination of a 
constant number of mean field many-body states at all times. In this respect  GCM$_R$ implementation is equivalent to many-configurations 
time-dependent mean field approximations~\cite{Yeager:1979,Haxton:2011,Wang:2015,Lode:2020}, 
when the many-body wave function is a linear combination of a finite and constant number $N(Q)$ of mean field many-body states.
The initial GCM$_R$  many-body wave function in case of induced fission with low energy neutrons has a relatively simple structure, since the nucleus
is physically a ``thermodynamically cold nucleus~\cite{Bohr:1956,Wheeler:1963}.'' The initial values of $f(Q,0)$ can be obtained from 
solving the corresponding stationary GCM$_G$ equation for the set of chosen initial values of $Q$. 
This type of time-dependent many-body wave function never experiences a ``fragmentation'' during the evolution of the system, 
in other words it never becomes more complex if the number $N(Q)$ is time independent.

A fissioning nucleus starting with a relatively simple intrinsic structure, 
as in induced fission with low energy neutrons, evolves in an energy region with an
increasing in time very high local energy level density, which increases  even past scission~\cite{Bulgac:2023c,Bulgac:2023,Bulgac:2024a}.
As \textcite{Bohr:1956,Wheeler:1963} discussed a long time ago, an assumptions which has proven correct the  analysis and interpretation of experimental data~\cite{Vandenbosch:1973,Wagemans:1991}, 
a fissioning nucleus near the outer fission barrier is ``thermodynamically cold''
and thus this initial state can be represented by a rather small number of generalized Slater determinants. During the slide from the top of outer fission barrier to the
scission configuration this nucleus ``heats up'' and the lower bound to the entropy of the system~\cite{Klich:2006,Bulgac:2023} is a very useful measure to evaluate
\begin{align}
S(t) = -\sumint_k [n_k(t)\ln n_k(t) + (1-n_k(t))\ln(1-n_k(t))], \label{eq:entropy}
\end{align}
where $n_k(t)$ are instantaneous occupation of single-particle states. The entropy $S(t)$  increases very fast with time 
$t$ in eGCM as this mechanism is intrinsically built into, 
and the single-particle occupation probabilities are continuously ``fragmented'' more and more
between many more generalized Slater determinants than at the initial time, 
a framework for which eGCM is ideally suited to describe. 

Another example would be the reaction discussed in Section~\ref{sec:V}, the collision $^{48}$Ca+$^{208}$Pb with the initial state 
$\Psi(\xi_1\ldots\xi_A,0)$ at $t=0$ of the two nuclei in their center of mass reference frame 
$(x,y,z)=(b\cos(\phi),b\sin(\phi),-\infty)$ with $\phi\in (0\ldots 2\pi)$ and $b\in(5\ldots 6)$ fm, which is a linear combination with equal weights of 
only $N({\bf b})=152$ Slater determinants.  In the final state at $t = \infty$ the wave function of the two reaction fragments 
will have components with various numbers of protons and neutrons and with their positions and momenta limited only by the conservation laws. 
This time-dependent eGCM many-body wave functions eGCM is a solution of the equations Eqs.~(\ref{eq:PsiT}-\ref{eq:eGCMTT})
 reproduced here for convenience in the same order
\begin{align}
&\Psi(\xi_1\ldots\xi_A,t)=\sumint_{k}h_{k}(t){\widetilde {\Phi}}_k(\xi_1\ldots\xi_A), \label{eq:PsiT1}\\
& \sumint_k\langle {\widetilde {\Phi}}_l|\hat{\rm H}|{\widetilde {\Phi}}_k\rangle h_{k}(t)=i\hbar\partial_t h_{l}(t),\label{eq:EGCM} \\
& h_k(0) = \int_{\xi_1\ldots\xi_A} {\widetilde {\Phi}}_k^*(\xi_1\ldots\xi_A) \Psi(\xi_1\ldots\xi_A,0),
\end{align}
with $\widetilde{\Phi}_k(\xi_1\ldots\xi_A)$ defined in Eq.~\eqref{eq:PsiQe}.

Both GCM$_R$ and eGCM simulations can be performed starting with a single (generalized) Slater determination wave function. The end results 
of these such simulations however will be drastically different.  In GCM$_R$ the final many-body wave functions will be a linear combination of at most 
$N({\bf b})$ mean field wave functions, while in eGCM the number of mean field wave functions in the final state, starting from the same initial states 
will be $N({\bf b},\tau)\gg N({\bf b})$.
Performing an eGCM  simulation with the same set of initial values of the impact parameter $N({\bf b})$ as in an GCM$_R$ simulation,
the corresponding eGCM final many-body wave functions is now significantly different than a GCM$_R$ many-body wave function, 
since now the final wave function is a linear combination  of 
\begin{align}
N({\bf b},\tau)\gg N({\bf b})\gg 1.
\end{align}  
Any single initial Slater determinant evolved under eGCM Eq.~\eqref{eq:EGCM} evolves into a many-body 
wave function with a much more complex structure, a  linear combination of $N({\bf b},\tau)$ (generalized) Slater determinants. 
This is a well known and well established result that if  the initial many-body state is chosen 
to be a mean field state, when propagated with the full many-body time-dependent Schr\"odinger 
equation the mean field character is lost in a very short time
and the time evolved many-body wave function with time becomes a linear combination of an increasing with time 
number of Slater determinants~\cite{Beylkin:2008}.
This is the fundamental difference between 
the existing GCM frameworks studied so far in literature and eGCM, 
with features obviously indeed required in a truly microscopic approach.

In order to obtain the time evolution of the entropy $S(t)$ defined in Eq.~\eqref{eq:entropy} one needs to determine the eigenvalues of 
the one-body density operator~\cite{Bulgac:2023} 
\begin{align}
&n(\xi,\zeta|t)=\langle\Psi(t) |\psi^\dagger(\zeta)\psi(\xi)|\Psi(t)\rangle,\\ 
&\int_\zeta n(\xi,\zeta | t)\phi_k(\zeta | t)=n_k(t)\phi_k(\xi | t), 
\end{align}
with $|\Psi(t)\rangle$ defined in either Eq.~\eqref{eq:GCM_R} in case of GCM$_R$ or in Eq.~\eqref{eq:PsiT1} in case of eGCM.  
As far as I am aware there are no studies in the literature 
on the time-dependent entropy $S(t)$ evaluated in approaches beyond TDDFT for nuclear reactions in general and fission in particular, an 
aspect of particular interest for Quantum Information Science~\cite{Calabrese:2005,Calabrese:2006,Alba:2017,Amico:2008,Horodecki:2009,Haque:2009,
Eisert:2010,Boguslawski:2014,Gigena:2015,Bengtsson:2017,Watrous:2018} and Quantum Computing in general 
and for the the Eigenstate Thermalization Hypothesis in 
particular~\cite{Neumann:1929,Neumann:2010,Goldstein:2010,Berry:1977,Berry:1991,Srednicki:1994,Bulgac:2024} .

The eGCM framework is expected to describe the structure of the correlated many-body wave functions, 
in particular interference and entanglement between different
``classical'' fission trajectories, parametrized in fission usually by the initial shapes $(Q_{20}, Q_{30})$ 
on the rim of the outer fission barrier. 
On physical grounds it is not fully clear at this time what other physically relevant many-body states were not accounted for within eGCM. 
The bending DoFs of the neck, not discussed here, are expected to be play an important role~\cite{Bulgac:2019d,Bulgac:2021,Scamps:2023a},
even though other authors have a different point of view~\cite{Randrup:2021,Vogt:2009,Vogt:2021}.

Since when approaching scission many TDDFT fission trajectories follow a very similar path, different 
trajectories originating at different initial points $Q$ will find themselves in relatively close proximity of each other, albeit each 
parametrized with a different running ``time'' $t$, which translates into to an actual FFs spatial separation. 
Having many different trajectories in close spatial proximity of each other will result into a strong mixing, 
as demonstrated in Section~ \ref{sec:V} and Fig.~\ref{fig:IPR}. 
eGCM is equally applicable to heavy-ion reactions and it
correctly restores the translational and azimuthal symmetry of the many-body wave functions and it can be used to restore 
other broken symmetries. Since LACM dissipation is now organically incorporated in TDDFT extended 
to pairing correlations, there seem to be no other limitation to achieve a detailed, and likely controlled,  microscopic 
approach to fission and to many nucleon transfer in nuclear reactions. Moreover, eGCM  very likely is a very strong candidate 
to produce a truly microscopic description of induced fission cross sections, which are notoriously known to be very hard to model. \\

{\bf Acknowledgements} \\

I am thankful to P.-G. Reinhard for reading a previous  version of the manuscript and making a number of very useful of suggestions.
I thank I. Stetcu, I. Abdurrahman, and M. Kafker for many discussions and comments, 
I. Abdurrahman for preparing Fig.~\ref{fig:DensityC}, M. Kafker for redrawing Fig.~\ref{fig:0} to my needs, 
for generating Fig.~ \ref{fig:ring},  for suggesting and evaluating IPR and generating  Fig.~ \ref{fig:IPR}, and for performing 
all the TDHF simulations for the reaction $^{48}$Ca+$^{208}$Pb discussed in 
Section~\ref{sec:V} on the Frontier supercomputer at ORNL using the code LISE~\cite{Shi:2020} and with detailed results 
in Ref.~\cite{Kafker:2026a}. 
I thank D. Vretenar for raising number of questions on the very first version of the manuscript (August, 2024). I am also thankful 
to the last referee for the vote of confidence that  eGCM project can be brought to a successful ending.
The funding from the Office of Science, Grant No. DE-FG02-97ER41014    
and also the partial support provided by NNSA cooperative Agreement DE-NA0003841 is greatly appreciated. 
This research used resources of the Oak Ridge Leadership Computing Facility, which is a U.S. DOE Office of
Science User Facility supported under Contract No. DE-AC05-00OR22725.

\providecommand{\selectlanguage}[1]{}
\renewcommand{\selectlanguage}[1]{}

\bibliography{local_fission}

\end{document}